\documentclass[10pt,letterpaper]{article}
\usepackage{opex3,cite}
\usepackage{graphicx}
\usepackage{dcolumn}
\usepackage{bm,amstext}
\usepackage[colorlinks=true,breaklinks=false,hypertex]{hyperref}
\usepackage[normalem]{ulem}
\usepackage{amsmath,amssymb}
\newcommand{\cs}{\chi^{(3)}_{\rm casc}}
\newcommand{\ns}{n_{2,\rm casc}^I}
\newcommand{\nk}{n_{2,\rm Kerr}^I}
\newcommand{\kp}{k^{(1)}}
\newcommand{\deff}{d_{\rm eff}}
\newcommand{\imm}{~{\rm mm^{-1}}}
\newcommand{\mic}{~\mu{\rm m}}
\begin{document}

\title{The anisotropic Kerr nonlinear refractive index of the beta-barium borate ($\beta$-{BaB}$_2${O}$_4$) nonlinear crystal}

\author{Morten Bache$^1$, Hairun Guo$^1$, Binbin Zhou$^1$, and Xianglong Zeng$^{1,2}$}
\address{$^1$DTU Fotonik, Department of Photonics Engineering, Technical University of Denmark, DK-2800 Kgs. Lyngby, Denmark\\
$^2$The Key Lab of Specialty Fiber Optics and Optical Access Network, Shanghai University, 200072 Shanghai, China}
\email{moba@fotonik.dtu.dk}\homepage{http://www.fotonik.dtu.dk/uno}
\begin{center}
(Complied \today)
\end{center}

\begin{abstract}
We study the anisotropic nature of the Kerr nonlinear response in a beta-barium borate ($\beta$-{BaB}$_2${O}$_4$, BBO) nonlinear crystal. The focus is on determining the relevant $\chi^{(3)}$ cubic tensor components that affect interaction of type I cascaded second-harmonic generation. Various experiments in the literature are analyzed and we correct the data from some of the experiments for contributions from cascading as well as for updated material parameters. We also perform an additional experimental measurement of the Kerr nonlinear tensor component responsible for self-phase modulation in cascading, and we show that the average value of 14 different measurements is considerably larger than what has been used to date. Our own measurements are consistent with this average value. We also treat data measurements for mixtures of tensor components, and by disentangling them we present for the first time a complete list that we propose as reference of the four major cubic tensor components in BBO. We finally discuss the impact of using the cubic anisotropic response in ultrafast cascading experiments in BBO.
\end{abstract}

\ocis{(190.3270) Kerr effect. (160.4330) Nonlinear optical materials. (190.4400) Nonlinear optics, materials. (190.7110) Ultrafast Optics. (190.5530) Pulse propagation and temporal solitons.}


\section{Introduction}
\label{sec:Introduction}

Cascaded second-harmonic generation (SHG) describes the case where the frequency-converted second harmonic (SH) is strongly phase mismatched, and this may create a Kerr-like nonlinear phase shift on the fundamental wave (FW). Since the early discovery of this cascaded self-action nonlinearity \cite{Ostrovskii:1967,Thomas:1972}, cascaded SHG was for a long time overlooked until DeSalvo et al. measured a negative nonlinear phase shift from phase-mismatched SHG using the Z-scan method \cite{desalvo:1992} (for an early comprehensive review on cascading, see \cite{stegeman:1996}). The exciting promise of cascading nonlinearities is seen directly from the scaling properties of the Kerr-like nonlinear index it generates: $\ns\propto-d_{\rm eff}^2/\Delta k$, where $d_{\rm eff}$ is the effective nonlinearity of the quadratic interaction. By simply changing the phase mismatch parameter $\Delta k$ we therefore have a Kerr-like nonlinearity that can be tuned in sign and strength. A particularly interesting and attractive property is that a self-defocusing Kerr-like nonlinearity $\ns<0$ is accessible by having a positive phase mismatch $\Delta k>0$.

A popular crystal for cascading experiments is beta-barium-borate ($\beta$-{BaB}$_2${O}$_4$, BBO), see e.g. \cite{liu:1999,Beckwitt:2001,ashihara:2002,ilday:2004,moses:2006,moses:2006b,moses:2007,bache:2010e,tan:1993,hache:1995,moses:2007a}. It has a decent quadratic nonlinear coefficient, and because the crystal is anisotropic it can be birefringence phase-matched for type I ($oo\rightarrow e$) SHG. For femtosecond experiments it has the important properties of a low dispersion, a high damage threshold, and a quite low Kerr self-focusing nonlinearity; the latter is important for cascading as we discuss later, because the material Kerr nonlinearity will compete with the induced cascading nonlinearity. The main goal of this paper is to analyze experiments in the literature where the Kerr nonlinear refractive index was measured in BBO \cite{tan:1993,hache:1995,desalvo:1996,sheik-bahae:1997,li:1997,Li:2001,ganeev:2003,Banks:2002,moses:2007a}, and for the fist time extract all four tensor components that are relevant to describe the anisotropic nature of the cubic nonlinearity.

While the anisotropy of the quadratic nonlinearity is extensively used for optimizing the phase-matching properties and maximizing the quadratic nonlinear coefficients in nonlinear crystals, the case is different when it comes to treating the cubic nonlinearities: the anisotropic nature has often been neglected when studying self-phase modulation (SPM) effects from the Kerr nonlinear refractive index, in contrast to the case of third-harmonic generation, as evidenced even in early studies \cite{Midwinter:1965,wang:1969}. Here we show which anisotropic nonlinear susceptibility components the experimental literature data represent. The experimental data will be analyzed and we correct the reported values for cascading contributions if necessary. We eventually obtain complete information of all four relevant tensor components for the BBO cubic nonlinear susceptibility. Our results show that the tensor component affecting the SPM of the ordinary wave, $c_{11}=\chi_{XXXX}^{(3)}=\chi_{YYYY}^{(3)}$, is very well documented, and the corrected literature data agree extremely well in the near-IR with the popular two-band model \cite{Sheik-Bahae:1991}. We also perform our own experimental measurement of $c_{11}$ with a very accurate technique based on balancing out the self-focusing SPM from the Kerr effect with the self-defocusing effect from cascading (originally used in \cite{moses:2007a}). Both this measurement and the corrected data from the literature show that $c_{11}$ is substantially larger than what has been used so far in simulations in the literature, and an important consequence for cascaded SHG is that this reduces the range of phase-mismatch values that gives a total defocusing nonlinearity. In contrast experiments measuring the tensor components that affect the extraordinary wave interaction are more scarce, so new measurements are necessary in order to get more accurate and reliable values. We also assess the impact of using the anisotropic Kerr response in simulations of cascaded SHG, and it turns out that for BBO the optimal interaction angles are such that the anisotropy plays a minor role in the wave dynamics.

As to analyze the experiments properly and rule out any misunderstandings in the definitions, the foundation of the paper is a careful formulation of the propagation equations of the FW and SH waves under slowly-varying envelope approximation, see appendix \ref{sec:SVEA}. These include the anisotropic nature of the frequency conversion crystal in formulating the effective cubic nonlinear coefficients. This theoretical background is important in order to analyze and understand the experiments from the literature that report Kerr nonlinearities for BBO. The analysis presented here should help understanding what exactly has been measured, and put the results into the context of cascaded quadratic soliton compression.

\section{Background for cascading nonlinearities}
\label{sec:cascading}

In order to understand the importance of knowing accurately the Kerr nonlinear index, let us explain first how cascading works. The cascaded Kerr-like nonlinearity can intuitively be understood from investigating the cascade of frequency-conversion steps that occur in a strongly phase-mismatched medium \cite{stegeman:1996}: After two coherence lengths $2\pi/|\Delta k|$ back-conversion to the FW is complete, and as $\Delta k\neq 0$ the SH has a different phase velocity than the FW. Thus, the back-converted FW photons has a different phase than the unconverted FW photons, simply because in the brief passage when they traveled as converted SH photons the nonzero phase mismatch implies that their phase velocities are different. For a strong phase mismatch $\Delta kL\gg 2\pi$, this process repeats many times during the nonlinear interaction length, and in this limit the FW therefore effectively experiences a nonlinear phase shift parameterized by a higher-order (cubic) nonlinear index $\ns$ given as \cite{desalvo:1992}
\begin{align}
    \label{eq:n2-SHG}
  \ns&=-\frac{2\omega_1 d_{\rm  eff}^2}{c^2\varepsilon_0 n_1^2n_2 \Delta k}
\end{align}
where $\omega_1$ is the FW frequency, $n_1$ and $n_2$ the linear refractive indices of the FW and SH, $\Delta k=k_2-2k_1$, and $k_j=n_j\omega_j/c$ are the wave vectors. The cascaded cubic nonlinearity will compete with the material cubic nonlinearity (the Kerr nonlinear index $\nk$), and the total nonlinear refractive index change experienced by the FW can essentially be described as
\begin{align}\label{eq:Deltan}
  \Delta n= n_{2,\rm tot}^{I}I_1=(\ns+\nk)I_1
\end{align}
where $I_1$ is the FW intensity. 
Due to the competing material Kerr nonlinearity a crucial requirement for a \textit{total }defocusing  nonlinearity $n_{2,\rm tot}^{I}<0$ is that the phase mismatch is low enough so that $|\ns|>\nk$ (the so-called Kerr limit). When this happens, cascaded SHG can generate strong spectral broadening through SPM on femtosecond pulses, which can be compensated in a dispersive element. In a pioneering experiment by Wise's group femtosecond pulse compression was achieved this way \cite{liu:1999}. Since the nonlinearity is self-defocusing the pulse energy is in theory unlimited as self-focusing effects are avoided \cite{liu:1999}, and it can even be used to heal small-scale and whole-scale self-focusing effects \cite{Beckwitt:2001}. Another very attractive feature of the negative self-defocusing nonlinearity is that temporal solitons can be excited in presence of normal (positive) group-velocity dispersion, which means anywhere in the visible and near-IR. With this approach Ashihara et al. \cite{ashihara:2002} achieved soliton compression of longer femtosecond pulses. Numerous experiments over the past decade have since been motivated by soliton compression of energetic pulses to few-cycle duration \cite{ashihara:2004,moses:2006,moses:2007,Zeng:2008,zhou:2012}. Since the defocusing limit and the soliton interaction depend critically on the $\nk$ value, knowing an exact value of the $\nk$ coefficient is crucial.

\section{Anisotropic quadratic and cubic nonlinearities in uniaxial crystals}
\label{sec:anisotropy}

\begin{figure}[t]
\begin{center}
\includegraphics[width=10cm]{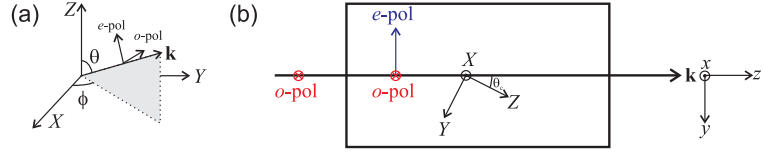}
\end{center}
\caption{\label{fig:cut} (a) The definition of the crystal coordinate system $XYZ$ relative to the beam propagation direction $\mathbf{k}$. 
(b) Top view of the optimal crystal cut for type I $oo\rightarrow e$ SHG in BBO, which has $\phi=-\pi/2$ and $\theta=\theta_c$ for perpendicular incidence of an $o$-polarized FW beam, and the $e$-polarized SH is generated through type I $oo\rightarrow e$ SHG. The specific value of the cut angle $\theta_c$ depends on the wavelength and the desired application. Angle-tuning the crystal in the paper plane will change the interaction angle $\theta$. Capital letters $XYZ$ are traditionally used to distinguish the crystal coordinate system from the beam coordinate system $xyz$ that has its origin in the $\mathbf{k}$-vector propagation direction.}
\end{figure}

In a uniaxial crystal, the isotropic base-plane is spanned by the crystal $XY$ axes, and light polarized in this plane is ordinary ($o$-polarized) and has the linear refractive index $n_o$. The optical axis (crystal $Z$-axis, also called the $c$-axis) lies perpendicular to this plane, and light polarized in this plane is extraordinary ($e$-polarized) and has the linear refractive index $n_e$. In a negative uniaxial crystal like BBO, $n_o>n_e$. The propagation vector ${\mathbf k}$ in this crystal coordinate system has the angle $\theta$ from the $Z$-axis and the angle $\phi$ relative to the $X$-axis, cf. Fig. \ref{fig:cut}(a), and the $e$-polarized component will therefore experience the refractive index $n^e(\theta)=[n_o^2/\cos^2\theta+n_e^2/\sin^2\theta]^{-1/2}$, while the $o$-polarized light always has the same refractive index.

Degenerate SHG can either be noncritical (type 0) interaction, where the FW and SH fields are polarized along the same direction, or critical (type I) interaction, where the FW and SH are cross-polarized along arbitrary directions in the $(\theta,\phi)$ parameter space. In type 0 the FW and SH are usually polarized along the crystal axes ($\theta=0$ or $\pi/2$) since this turns out to maximize the nonlinearity, and it is called noncritical because the interaction does not depend critically on the propagation angles (both nonlinearity and phase matching parameters vary little with angle). A great advantage of this interaction is that there is little or no spatial walk off (a consequence of the choice $\theta=0$ or $\pi/2$). In type I $\theta$ and $\phi$ values can often be found where phase matching is achieved. This interaction is very angle-sensitive, which is why it is called critical interaction.

The slowly-varying envelope equations (SVEA) for degenerate SHG are derived in Appendix \ref{sec:SVEA}. There the quadratic and cubic "effective" anisotropic nonlinear coefficients were symbolically introduced. Below we show the expressions for these coefficients relevant for SHG in BBO, which belongs to the crystal class $3m$, and in all cases we consider the reduced numbers of tensor components that come from adopting Kleinman symmetry (which assumes dispersionless nonlinear susceptibilities \cite[Ch. 1.5]{boyd:2007}); thus BBO has 3 independent $\chi^{(2)}$ tensor components and 4 independent $\chi^{(3)}$ tensor components. An excellent overview of the quadratic and cubic nonlinear coefficients in other anisotropic nonlinear crystal classes is found in \cite{Banks:2002}.

For an arbitrary input (i.e. FW) polarization various SHG processes can come into play ($oo\rightarrow o$, $oo\rightarrow e$, $oe\rightarrow e$, $oe\rightarrow o$, $ee\rightarrow e$, and $ee\rightarrow o$). This includes both type 0, type I and type II (nondegenerate SHG, where the FW photons are cross-polarized). Their respective $\deff$-values  for a crystal in the $3m$ point group (which also includes lithium niobate) are \cite{Banks:2002}
\begin{align}
\label{eq:deff-ooo}  \deff^{ooo}&=-d_{22}\cos3\phi\\
\label{eq:deff-ooe}  \deff^{ooe}&=\deff^{oeo}=d_{31}\sin(\theta+\rho)-d_{22}\cos(\theta+\rho)\sin3\phi\\
\label{eq:deff-eeo}  \deff^{oee}&=\deff^{eeo}=d_{22}\cos^2(\theta+\rho)\cos 3\phi\\
\label{eq:deff-eee}  \deff^{eee}&=d_{22}\cos^3(\theta+\rho)\sin3\phi +3d_{31}\sin(\theta+\rho)\cos^2(\theta+\rho)+d_{33}\sin^3(\theta+\rho)
\end{align}
We remark here that the correct evaluation of the effective nonlinearity requires to take into account the spatial walk-off angle $\rho=\arctan [\tan(\theta)n_o^2/n_e^2]-\theta$ (for a negative uniaxial crystal) \cite{Nikogosyan:1991}. In BBO $\rho$ is around $3-4^\circ$ for the typical angles used, which gives changes in $d_{\rm eff}$ of a few percent, and as this is well within the standard errors on the experimental $d_{ij}$ values it is typically ignored for qualitative simulations. However, when we later evaluate the cascading contributions and compare them to the competing material Kerr nonlinearity, an exclusion of $\rho$ would lead to a systematic error so we choose to include it in the analysis in Sec. \ref{sec:cubic}.

BBO is usually pumped with $o$-polarized light as the quadratic nonlinearity is largest for this configuration ($d_{22}$ is the largest tensor component, see App. \ref{sec:Crystal-parameters}), and because the $oo\rightarrow e$ interaction can be phase matched by a suitable angle $\theta$. For such an interaction, the crystal is cut so $\phi=-\pi/2$ as to optimize the nonlinearity (it is here relevant to mention that $d_{22}$ and $d_{31}$ has opposite signs in BBO). This has the direct consequence that $oo\rightarrow o$ interaction is zero. For this reason, when studying BBO pumped with $o$-polarized light in a crystal with $\phi=-\pi/2$, Eq. (\ref{eq:deff-ooe}) shows that $d_{\rm eff}=d_{31}\sin(\theta+\rho) -d_{22}\cos(\theta+\rho)$.

\label{p:BBO-3m}
Note that BBO has historically been misplaced in the point group 3, and in addition there has been some confusion about the assignment of the crystal axes (where the mirror plane of the crystal was taken parallel instead of perpendicular to the crystal $X$-axis) \cite{Nikogosyan:1991}. Unfortunately this means that even today crystal company web sites operate with $d_{11}$ as being the largest tensor component (in the $3m$ point group $d_{11}=0$) and supply crystals apparently cut with $\phi=0$ (because using the point group 3 combined with a nonstandard crystal axes definition means that $d_{11}\cos 3\phi$ must be maximized, while with the correct point group, $3m$, and correct crystal axes assignments, $d_{22}$ is the largest tensor component and $\sin 3\phi=1$ maximizes the effective nonlinearity). In order to sort out any confusion, the optimal crystal cut is shown in Fig. \ref{fig:cut}(b), and we hereby urge crystal companies to follow the standard notation.

In Appendix \ref{sec:SVEA} the SVEA propagation equations included also an "effective" third-order SPM nonlinearity, $\chi^{(3)}_{\rm eff}(\omega_j;\omega_j)$, and cross-phase modulation (XPM) nonlinearity, $\chi^{(3)}_{\rm eff}(\omega_j;\omega_k)$, which due to the anisotropy had to be calculated specifically for a given crystal class and input polarization. The SPM and XPM anisotropic cubic nonlinearities for a uniaxial crystal in the point group $3m$ were found in Eqs. (B13), (B14) and (B15) in \cite{bache:2010}. For a type I interaction ($oo\rightarrow e$), where the FW is $o$-polarized and the SH $e$-polarized, they are
\begin{align}\label{eq:chi3-SPM-FW}
    \chi^{(3)}_{\rm eff}(\omega_1;\omega_1)=&c_{11}\\
  \chi^{(3)}_{\rm eff}(\omega_2;\omega_2)=&-4c_{10}\sin(\theta+\rho)\cos^3(\theta+\rho)\sin 3\phi+  c_{11}\cos^4(\theta+\rho) \nonumber \\&
  +\tfrac{3}{2}c_{16}\sin^2(2\theta+2\rho)+c_{33}\sin^4(\theta+\rho)
\label{eq:chi3-SPM-SH}
 \\
\label{eq:chi3-XPM}   \chi^{(3)}_{\rm eff}(\omega_1;\omega_2)=&\tfrac{1}{3}c_{11}\cos^2(\theta+\rho)
 +c_{16}\sin^2(\theta+\rho) +c_{10}\sin (2\theta+2\rho)\sin 3\phi
\end{align}
Instead for a type 0 $ee\rightarrow e$ interaction, as recently considered for cascading quadratic nonlinearities in lithium niobate \cite{zhou:2012} and periodically poled lithium niobate \cite{Langrock:2007,Phillips:2011}, we have
\begin{align}\label{eq:chi3-type0-general}
    \chi^{(3)}_{\rm eff}(\omega_1;\omega_1)&=\chi^{(3)}_{\rm eff}(\omega_2;\omega_2)=\chi^{(3)}_{\rm eff}(\omega_1;\omega_2)
   \\
   =&-4c_{10}\sin(\theta+\rho)\cos^3(\theta+\rho)\sin 3\phi+
  c_{11}\cos^4(\theta+\rho) \nonumber \\&
  +\tfrac{3}{2}c_{16}\sin^2(2\theta+2\rho)+c_{33}\sin^4(\theta+\rho)
\label{eq:chi3-type0}
\end{align}
We have here used the contracted notation $c_{\mu m}\equiv \chi_{ijkl}^{(3)}$, where the indices $i,j,k,l$ can take the values $X,Y,Z$ and \cite{boulanger:2006}
\begin{align}
  {\rm for~\mu:} &\quad X \rightarrow 1\quad Y \rightarrow 2
  \quad Z \rightarrow 3\nonumber\\
  {\rm for~}m: &\quad XXX \rightarrow 1 \quad YYY \rightarrow 2
  \quad  ZZZ \rightarrow 3 \quad YZZ \rightarrow 4
  \quad YYZ \rightarrow 5 \nonumber\\& \quad  XZZ \rightarrow 6 \quad XXZ \rightarrow 7
  \quad XYY \rightarrow 8 
  \quad XXY \rightarrow 9 \quad XYZ  \rightarrow 0
\end{align}
Note that by using Kleinman symmetry it is assumed that the chromatic dispersion of the nonlinearities is negligible. 
However, identities like $\chi^{(3)}_{\rm eff}(\omega_1;\omega_1)= \chi^{(3)}_{\rm eff}(\omega_2;\omega_2)=\chi^{(3)}_{\rm eff}(\omega_1;\omega_2)$ as expressed by Eq. (\ref{eq:chi3-type0-general}) have empirically been found not to hold. Specifically, the tensor components they are calculated from turn out to obey slight frequency variations that can be predicted by frequency scaling rules, like Miller's rule (see \cite{miller:1964,wynne:1969,Ettoumi:2010} and also later in Sec. \ref{sec:cij_summary}).

If the Kerr nonlinearity is considered isotropic, we have $\chi_{\rm eff}^{(3)}(\omega_1;\omega_2)=\chi_{{\rm eff}}^{(3)}(\omega_1;\omega_1)/3=\chi_{{\rm eff}}^{(3)}(\omega_2;\omega_2)/3$ [type I $oo\rightarrow e$ interaction, taking $\theta=0$ in Eqs. (\ref{eq:chi3-SPM-FW})-(\ref{eq:chi3-XPM})], or $\chi_{{\rm eff}}^{(3)}(\omega_1;\omega_1)=\chi_{{\rm eff}}^{(3)}(\omega_2;\omega_2)=\chi_{{\rm eff}}^{(3)}(\omega_1;\omega_2)$ [type 0 $ee\rightarrow e$ interaction, cf. Eq. (\ref{eq:chi3-type0-general})]. These properties underlie the parameter $B$ that we used in our previous isotropic model \cite{bache:2007}.

A mistake often seen in the early literature on type I interaction is to take $\chi_{\rm SPM}^{(3)}=\chi_{\rm XPM}^{(3)}$, and this mistake gives a 3 times too large XPM term. As shown above the identity $\chi_{\rm SPM}^{(3)}=\chi_{\rm XPM}^{(3)}$ only holds in the type 0 configuration, where the FW and SH have identical polarizations, but importantly it is not an identity that is restricted to isotropic nonlinearities as it holds for an anisotropic medium as well. Thus, the error made in the past for type I  could either come from using directly the propagation equations for an isotropic medium and where two pulses with the same polarization interact, but it could also come from generalizing type 0 SHG propagation equations to type I, and forgetting the XPM properties for cross-polarized interaction. 

\section{Measurement of cubic nonlinearities of BBO}
\label{sec:cubic}


\subsection{Experimental conditions for the literature measurements}
\label{sec:remarks}

Most experiments aiming to measure the cubic nonlinearities have used the Z-scan method \cite{sheik-bahae:1990}, which uses a Gaussian laser beam in a tight-focus limiting geometry to measure the Kerr nonlinear refractive index. The Z-scan method measures the transmittance of a nonlinear medium passing through a finite aperture placed in the far field as a function of the sample position ($z$) measured with respect to the focal plane. As the sample is Z-scanned (i.e. translated) through the focus of the beam, the lens effect from the Kerr nonlinear index change will change the amount of light recorded by the detector; this gives information about the intensity-induced nonlinear index change $\Delta n$ and thus the $n_2^I$ coefficient, defined phenomenologically from the general expansion
\begin{align}\label{eq:Deltan_n2}
    \Delta n=n_2^I I+n_4^I I^2+\cdots
\end{align}
The intensity is rarely high enough to allow for any contributions but the $n_2^I$ term. The superscript $I$ underlines that the $n_2^I$ parameter is the intensity-dependent nonlinear index, as one can also define an electric-field dependent index, typically as $\Delta n=\tfrac{1}{2}n_2^E |E|^2$. In the other sections of this paper we use $n_{2,\rm Kerr}^I$ to denote the material Kerr nonlinear refractive index and $n_{2,\rm casc}^I$ the Kerr-like cascading nonlinear index from cascaded SHG; this notation should suffice to avoid confusion with the linear refractive index of the SH, $n_2$. As the result of a closed-aperture measurement can be influenced by  contributions from multi-photon absorption, the aperture can be removed (open aperture scan). The $n_2^I$ contributions then vanish, and only multi-photon absorption effects remain. Thus it is possible to separate the two contributions.

There are some issues with the Z-scan method that may affect the measured nonlinear refractive index. If the repetition rate is too high, there are contributions to the measured $n_2^I$ from thermal effects as well as two-photon excited free carriers \cite{krauss:1994} (for more on these issues, see e.g. \cite{Gnoli:2005}). Similarly, a long pulse duration can also lead to more contributions to the measured $\nk$ than just the electronic response that we aim to model, and in particular the static (DC) Raman contribution is measured as well. However, for BBO the fraction of the delayed Raman effects is believed to be quite small, so we will assume that it can be neglected making the measured nonlinearities correspond to the electronic response.
In the following we will analyze measurements in the literature in order to extract the four BBO cubic tensor components in Eqs.~(\ref{eq:chi3-SPM-FW})-(\ref{eq:chi3-XPM}). 
When converting from the nonlinear susceptibilities $c_{\mu m}$ to the intensity nonlinear refractive index we can use Eq. (\ref{eq:n2-chi3}). There is a small caveat here, because the nonlinear index is defined for SPM and XPM terms, where only two waves interact through their intensities, while the $c_{\mu m}$ coefficients are generally defined for four-wave mixing. Thus, it does not always make sense to define a nonlinear refractive index (consider e.g. the $c_{10}$ component) until the total effective susceptibility is calculated, e.g. through Eqs. (\ref{eq:chi3-SPM-FW})-(\ref{eq:chi3-type0}). Thus, it is safer to keep the susceptibility notation as long as possible when calculating the effective nonlinearities. The exception is when measuring the nonlinear refractive index of the $oooo$ SPM interaction because the relevant nonlinear tensor component that is measured is $c_{11}$, see Eq. (\ref{eq:chi3-SPM-FW}), and thus the connection between $c_{11}$ (i.e. $\chi_{oooo}^{(3)}$) and $\nk$ is unambiguous.


In the experiments in the literature the cascaded quadratic contributions were often forgotten or assumed negligible. The exact relations for the evaluating cascading nonlinearities are derived in Appendix \ref{sec:NLSE}. We will address these issues below by in each case calculating where possible the cascading nonlinearity.

\subsection{The $c_{11}$ tensor component}
\label{sec:c11}

In this section we present measurements using $o$-polarized input light, which means that the $c_{11}$ tensor component is accessed by measuring the SPM component of the $o$-polarized light $\chi^{(3)}_{\rm eff}(\omega_1;\omega_1)=\chi^{(3)}_{oooo}(-\omega_1;\omega_1,-\omega_1,\omega_1)$. The connection to the Kerr nonlinear index is $n_{2,\rm Kerr}^I(\omega_1;\omega_1)=3\chi^{(3)}_{\rm eff}(\omega_1;\omega_1)/4n_1^2 \varepsilon_0 c=3c_{11}/4n_1^2 \varepsilon_0 c$. 

To our knowledge, the first measurement of any of the cubic susceptibilities was performed by Tan et al. \cite{tan:1993}. They investigated the cascaded nonlinearity in a 10 mm BBO crystal cut at $\theta=22.8^\circ$ for phase matching at 1064 nm (they claim also that the crystal is cut with $\phi=0$, which is probably due to the confusion related to placing BBO in the point group 3 instead of $3m$, cf. the discussion on p. \pageref{p:BBO-3m}). Pumping with 30 ps $o$-polarized pulses from a Nd:YAG mode-locked laser at two nearly degenerate wavelengths (a strong wave at $\lambda_a=1.064\mic$ and a weaker wave at $\lambda_b=1.090\mic$), they rotated the BBO crystal around $\Delta k=0$ in a type I $oo\rightarrow e$ SHG setup, where both the SH $2\omega_a$ of the strong beam was generated and four-wave mixing of the two pumps $\omega_c=2\omega_a-\omega_b$ was generated. By recording the impinging and generated pulse energies vs. angle they found by a fitting analysis the coefficient $g=220$, where $g=\bar\chi^{(3)}_{\rm int}/d_{\rm eff}^2$ was the only free parameter in the fit. For $\theta=22.8^\circ$ we have $\rho=3.2^\circ$ and $d_{\rm eff}=1.99$ pm/V, so this gives the measured nonlinear susceptibility $\bar\chi^{(3)}_{\rm int}=8.8\times 10^{-22}~{\rm m^2/V^2}$. They also performed a Z-scan measurement, and the setup was calibrated towards a BK7 glass sample. They found $\bar\chi^{(3)}_{\rm int}/\bar\chi^{(3)}_{\rm BK7}\simeq 1.4$. By using the value they propose $\bar\chi^{(3)}_{\rm BK7}=4.5\times 10^{-22}~{\rm m^2/V^2}$, one therefore finds $\bar\chi^{(3)}_{\rm int}=6.4\times 10^{-22}~{\rm m^2/V^2}$. We report these results with a bar because their definition of the cubic susceptibility is different than what we use: they define the SPM polarization as $\mathcal{P}_{\rm NL}^{(3)}=\varepsilon_0\bar\chi_{\rm int}^{(3)}\mathcal{E}|\mathcal{E}|^2/2$, which replaces our definition Eq. (\ref{eq:PNL3}), so the connection between them is $\bar\chi_{\rm int}^{(3)}=3/2 \chi_{\rm eff}^{(3)}(\omega_1;\omega_1)$. To check this, from the literature we find for BK7 the value $n_{2, \rm Kerr}^I=3.75\pm 0.3\cdot 10^{-20} ~{\rm m^2/W}$ measured at 804 nm by Nibbering et al. \cite{Nibbering:1995}, which corresponds to $\chi^{(3)}=3.03\times 10^{-22}~{\rm m^2/V^2}$. This is precisely a factor 3/2 smaller than the value Tan et al. mentions, and this confirms the relationship between the cubic susceptibilities. We now apply Miller's rule to the Nibbering et al. result to get for BK7 $\chi^{(3)}=2.90\times 10^{-22}~{\rm m^2/V^2}$ at 1064 nm, so we get a corrected value $\bar\chi^{(3)}_{\rm int}=6.08\times 10^{-22}~{\rm m^2/V^2}$. In our notation their cascading measurements gave
\begin{align}\label{eq:Tan-c11}
    c_{11}&=5.84\pm 0.51\cdot 10^{-22} ~{\rm m^2/V^2}, \; \lambda=1.064\mic~ 
\end{align}
corresponding to $\nk(\omega_1;\omega_1)=6.03\pm 0.52\cdot 10^{-20} ~{\rm m^2/W}$. Their Z-scan measurement gives
\begin{align}\label{eq:Tan-c11-Z}
    c_{11}&=4.05\pm 0.52\cdot 10^{-22} ~{\rm m^2/V^2}, \; \lambda=1.064\mic~ 
\end{align}
corresponding to $\nk(\omega_1;\omega_1)=4.18\pm 0.54\cdot 10^{-20} ~{\rm m^2/W}$. The error bars are indicative as they do not report the errors in their measurements, but we assume 5\% error on $g$ and conservatively estimate an error of 10\% on the Z-scan result. A question remains concerning the Z-scan result: they do not report the angle $\theta$ at which the measurement was done so we cannot asses the amount of cascading contributions they had in this measurement, but given the fact that the $c_{11}$ value is quite small it seems likely that cascading had a contribution.

Hache et al. \cite{hache:1995} used a 1 mm BBO cut at $\theta_c=29.2^\circ$ with 800 nm 100 fs pulses from a MHz-repetition rate Ti:sapphire oscillator. The pump was $o$-polarized allowing for $oo\rightarrow e$ cascading interaction. They carried out Z-scan measurements both close to the phase-matching point $\theta_c$ as well as "far from SHG phase matching". In the latter case they found 
\begin{align}\label{eq:Hache-nKerr11}
    \nk(\omega_1;\omega_1)=4.5\pm 1.0 \cdot 10^{-20} ~{\rm m^2/W}, \; \lambda=0.8\mic
\end{align}
which corresponds to $c_{11}=4.36\pm 0.97\cdot 10^{-22} ~{\rm m^2/V^2}$. They claim that cascading contributions did not contribute for this measurement, but whether this is true cannot be judged with the information at hand, as they did not specify the angle they used for this measurement.

DeSalvo et al. \cite{desalvo:1996} used the Z-scan method with a single-shot 30 ps pulse at 1064 nm, which had ${\mathbf k}$ parallel to the optical axis $c$, so  $\theta=0$ and consequently $\rho=0$ as well. Thus, (a) the pump is always $o$-polarized no matter how the input beam is polarized, and (b) the input polarization determines the angle $\phi$. Since the specific orientation of the crystal with respect to the input polarization was not reported, we take it as unknown. The relevant Kerr contribution at the pump wavelength is $oooo$ interaction, which is given by Eq. (\ref{eq:chi3-SPM-FW}) as simply one tensor component $c_{11}$. The experiment reported $n_{2}^E=11\pm 2\cdot 10^{-14}$ esu at 1064 nm, where the nonlinear index change is defined as $\Delta n=\tfrac{1}{2}n_{2}^E |E|^2$, and using Eq. (C1) in \cite{bache:2010} this corresponds to $\chi^{(3)}=2.70\pm 0.49\cdot 10^{-22} ~{\rm m^2/V^2}$. 
The cascading contributions are from $oo\rightarrow e$ and $oo\rightarrow o$ processes. Since $\theta=0$ they have the same phase mismatch values as $e$-polarized light in this case has the same refractive index as $o$-polarized light. At 1064 nm the value is $\Delta k_{ooe}=\Delta k_{ooo}=234\imm$. Taking into account the two cascading channels Eqs. (\ref{eq:deff-ooe}) and (\ref{eq:deff-ooo}), and using Eq. (\ref{eq:chi3-SHG}), we get the total contribution $\cs=-[\sin^2(3\phi)+\cos^2(3\phi)]1.95\cdot 10^{-22}~{\rm m^2/V^2}=-1.95\cdot 10^{-22}~{\rm m^2/V^2}$, i.e. independent on the propagation angle $\phi$.
For these calculations we used $d_{22}=-2.2$ pm/V. The $d_{\rm eff}$ values are typically reported with a 5\% uncertainty \cite{Shoji:1999}, giving a 7\% uncertainty on the cascading estimate. This means that when correcting for the cascading contributions we get 
\begin{align}\label{eq:deSalvo-c11}
    c_{11}&=4.65\pm 0.51\cdot 10^{-22} ~{\rm m^2/V^2}, \; \lambda=1.064\mic 
\end{align}
corresponding to $\nk(\omega_1;\omega_1)=4.80\pm 0.53\cdot 10^{-20} ~{\rm m^2/W}$. At 532 nm $n_2^E=21\pm 4\cdot 10^{-14}$ esu was measured, corresponding to  $\chi^{(3)}=5.22\pm 0.99\cdot 10^{-20} ~{\rm m^2/V^2}$. The cascading is smaller here, using $\Delta k_{ooe}=\Delta k_{ooo}=1,933\imm$ we get $\cs=-0.61\cdot 10^{-22}~{\rm m^2/V^2}$ (where we used $d_{22}=2.6$ pm/V as measured at 532 nm \cite{Shoji:1999}), so
\begin{align}\label{eq:deSalvo-c11-532}
    c_{11}&=5.82\pm 0.99\cdot 10^{-22} ~{\rm m^2/V^2}, \; \lambda=0.532\mic 
\end{align}
corresponding to $\nk(\omega_1;\omega_1)=5.87\pm 1.00\cdot 10^{-20} ~{\rm m^2/W}$. At 355 nm $n_{2}^E=14\pm 3\cdot 10^{-14}$ esu was measured, corresponding to  $\chi^{(3)}=3.54\pm 0.67\cdot 10^{-22} ~{\rm m^2/V^2}$. The cascading is small, using $\Delta k_{ooe}=\Delta k_{ooo}=11,736\imm$ we get $\cs=-0.38\cdot 10^{-22}~{\rm m^2/V^2}$ (where Miller's scaling was used to get the value $d_{22}=-4.4$ pm/V), so
\begin{align}\label{eq:deSalvo-nKerr11-355}
    c_{11}&=3.92\pm 0.68\cdot 10^{-22} ~{\rm m^2/V^2}, \; \lambda=0.355 \mic 
\end{align}
corresponding to $\nk(\omega_1;\omega_1)=3.81\pm 0.74\cdot 10^{-20} ~{\rm m^2/W}$. We must mention that the cascading value is quite uncertain: the SH wavelength is here 177.5 nm, which is very close to the UV poles in the Sellmeier equations (the fit in \cite{Zhang:2000} puts the poles at 135 nm and 129 nm for $o$- and $e$-polarized light, respectively). Finally, at 266 nm they measured $n_2^E=1\pm 0.3\cdot 10^{-14}$ esu. Here cascading is estimated to be insignificant (certainly we cannot estimate accurately it using the Sellmeier equations as the SH lies right at the UV pole), so converting to SI we have
\begin{align}\label{eq:deSalvo-c11-266}
    c_{11}&=0.26\pm 0.078\cdot 10^{-22} ~{\rm m^2/V^2}, \; \lambda=0.266\mic
\end{align}
corresponding to $\nk(\omega_1;\omega_1)=0.24\pm 0.07\cdot 10^{-20} ~{\rm m^2/W}$.

Li et al. \cite{li:1997,Li:2001} performed two different Z-scan experiments with similar conditions:  a thin $Z$-cut BBO crystal was pumped with 25 ps 10 Hz pulses at 532 nm and 1064 nm \cite{li:1997} and later with 150 fs 780 nm pulses from a 76 MHz Ti:sapphire oscillator \cite{Li:2001}. The propagation was along the optical axis, implying $\theta=0$ and consequently $\rho=0$ as well, and also that for both cascading and Kerr nonlinearities the angle $\phi$ should not matter as the pump always will be $o$-polarized. Nonetheless, they measured using the Z-scan method different results with the input polarization along the [1 0 0] direction ($X$-axis, i.e. $\phi=0$), and with input polarization along the [0 1 0] direction ($Y$-axis, i.e.  $\phi=\pi/2$). Cascading was in both papers ignored as it was considered too small, so let us assess whether this is true. The cascading contributions are from $oo\rightarrow e$ and  $oo\rightarrow o$, both having the same $\Delta k_{ooe}=\Delta k_{ooo}$. However, only one come into play in each case: for $\phi=0$ the only nonzero nonlinearity is $d_{{\rm eff}}^{ooo}=-d_{22}$ while for $\phi=\pi/2$ the only nonzero nonlinearity is $d_{{\rm eff}}^{ooe}=-d_{22}$. Thus, they turn out to have the same cascading contribution, which from Eq. (\ref{eq:n2-SHG}) becomes $\ns=-0.61\cdot 10^{-20}~{\rm m^2/W}$ (532 nm), $\ns=-1.30\cdot 10^{-20}~{\rm m^2/W}$ (780 nm), and $\ns=-2.01\cdot 10^{-20}~{\rm m^2/W}$ (1064 nm). When correcting the measured nonlinearities with this value we get for the picosecond experiments at 532 nm \cite{li:1997}
\begin{align}
\label{eq:Li-nKerr11-x-1997}
    \nk(\omega_1;\omega_1)=&5.41\pm 0.90\cdot 10^{-20} ~{\rm m^2/W}, 
    \quad [1 \;0\; 0],\; \lambda=0.532\mic
\\
\label{eq:Li-nKerr11-y-1997}
    \nk(\omega_1;\omega_1)=&4.61\pm 0.80\cdot 10^{-20} ~{\rm m^2/W}, 
    \quad [0 \;1\; 0],\; \lambda=0.532\mic
\end{align}
corresponding to $c_{11}=5.37\pm 0.89\cdot 10^{-22} ~{\rm m^2/V^2}$ and $c_{11}=4.58\pm 0.79\cdot 10^{-22} ~{\rm m^2/V^2}$, respectively. At 1064 nm we get
  \begin{align}
\label{eq:Li-nKerr11-1997-1064nm}
    \nk(\omega_1;\omega_1)=&7.01\pm 1.01\cdot 10^{-20} ~{\rm m^2/W}, 
\quad \lambda=1.064\mic
\end{align}
corresponding to $c_{11}=6.79\pm 0.98\cdot 10^{-22} ~{\rm m^2/V^2}$. For this measurement the polarization direction is unknown. Instead for the femtosecond experiments we get \cite{Li:2001}
\begin{align}
\label{eq:Li-nKerr11-x}
    \nk(\omega_1;\omega_1)=&5.30\pm 0.51\cdot 10^{-20} ~{\rm m^2/W}, 
    \quad [1 \;0\; 0],\; \lambda=0.78\mic
\\
\label{eq:Li-nKerr11-y}
    \nk(\omega_1;\omega_1)=&4.50\pm 0.51\cdot 10^{-20} ~{\rm m^2/W}, 
    \quad [0 \;1\; 0],\; \lambda=0.78\mic
\end{align}
corresponding to $c_{11}=5.18\pm 0.50\cdot 10^{-22} ~{\rm m^2/V^2}$ and $c_{11}=4.39\pm 0.50\cdot 10^{-22} ~{\rm m^2/V^2}$, respectively. When reducing the repetition rate from 76 MHz to 760 kHz they saw similar results.

Ganeev et al. \cite{ganeev:2003} used a 2 Hz 55 ps 1064 nm pump pulse and a BBO crystal with $\theta=51^\circ$ and consequently $\rho=4.1^\circ$. The procedure they used was to calculate the "critical" $\Delta k_c$ where defocusing cascading exactly balances the intrinsic focusing Kerr, i.e. where $|\ns|=\nk$. By angle tuning well away from this point (so $\Delta k\gg\Delta k_c$) they tried to make cascading insignificant. We estimate this to be true: specifically, the pump was $o$-polarized and the crystal was cut with $\phi=-\pi/2$ to optimize SHG \cite{Ganeev:2012}. Then the only cascading channel is $oo\rightarrow e$, with $\Delta k_{ooe}=-655\imm$, and using Eq. (\ref{eq:n2-SHG}) we get $\ns=0.26\cdot 10^{-20}~{\rm m^2/W}$ that is therefore self-focusing (as they also note themselves). Using the Z-scan method they measured $n_2^I=7.4\pm 2.2\cdot 10^{-20} ~{\rm m^2/W}$ with the uncertainty estimated to be $30\%$, so correcting for cascading
\begin{align}\label{eq:ganeev-nKerr11-1064}
    \nk(\omega_1;\omega_1)=7.14\pm 2.22\cdot 10^{-20} ~{\rm m^2/W}, \; \lambda=1.064\mic
\end{align}
corresponding to $c_{11}=6.92\pm 2.15\cdot 10^{-22} ~{\rm m^2/V^2}$. At 532 nm they measured $n_2^I=8.0\pm 2.4\cdot 10^{-20} ~{\rm m^2/W}$. Here cascading is more significant as $\Delta k_{ooe}=-194\imm$, $\rho=4.8^\circ$, $d_{\rm eff}=1.50$ pm/V and $\ns=2.22\cdot 10^{-20}~{\rm m^2/W}$. Therefore the Kerr value corrected for cascading becomes
\begin{align}\label{eq:ganeev-nKerr11-532}
    \nk(\omega_1;\omega_1)=5.24\pm 2.41\cdot 10^{-20} ~{\rm m^2/W}, \; \lambda=0.532\mic
\end{align}
which corresponds to $c_{11}=5.78\pm 2.41\cdot 10^{-22} ~{\rm m^2/V^2}$.

Moses et al. \cite{moses:2007a} pumped a BBO crystal at 800 nm with 110 fs $o$-polarized light from a 1 kHz Ti:Sapphire regenerative amplifier. By angle-tuning the crystal they achieved zero nonlinear refraction: this happens when the cascading from the $oo\rightarrow e$ interaction exactly balances the Kerr nonlinear refraction. Investigating a range of intensities they determined this \textit{ critical} phase mismatch value to $\Delta k_c=31\pm 5~\pi/\rm mm$. Using this value and assuming that it corresponds to $n_{2,\rm tot}^I=0$ then a reverse calculation though Eq. (\ref{eq:n2-SHG}) gives $\nk=-\ns$. Using the value $d_{\rm eff}=1.8$ pm/V they inferred the value $\nk=4.6\pm 0.9\cdot 10^{-20} ~{\rm m^2/W}$ (the main sources of the uncertainty are to determine the precise phase-mismatch value as well as the uncertainty on $d_{\rm eff}$). Since $o$-polarized pump light was used, this contains only the $c_{11}$ tensor contribution from an $oooo$ interaction. However, the value $d_{\rm eff}=1.8$ pm/V they used to infer this nonlinearity was probably taken too low \cite{moses:2010comm}. Let us therefore estimate it again using a more accurate value: $\Delta k=31~\pi/\rm mm$ is achieved at 800 nm with $\theta=26.0^\circ$, giving $\rho=3.6^\circ$ and $d_{\rm eff}=2.05$ pm/V. With this corrected value we get
\begin{align}\label{eq:Moses-nKerr11}
    \nk(\omega_1;\omega_1)=5.54\pm 0.98 \cdot 10^{-20} ~{\rm m^2/W}, \; \lambda=0.8\mic
\end{align}
which corresponds to $c_{11}=5.41\pm 0.95\cdot 10^{-22} ~{\rm m^2/V^2}$.

\begin{figure}[t]
\begin{center}
\includegraphics[width=6cm]{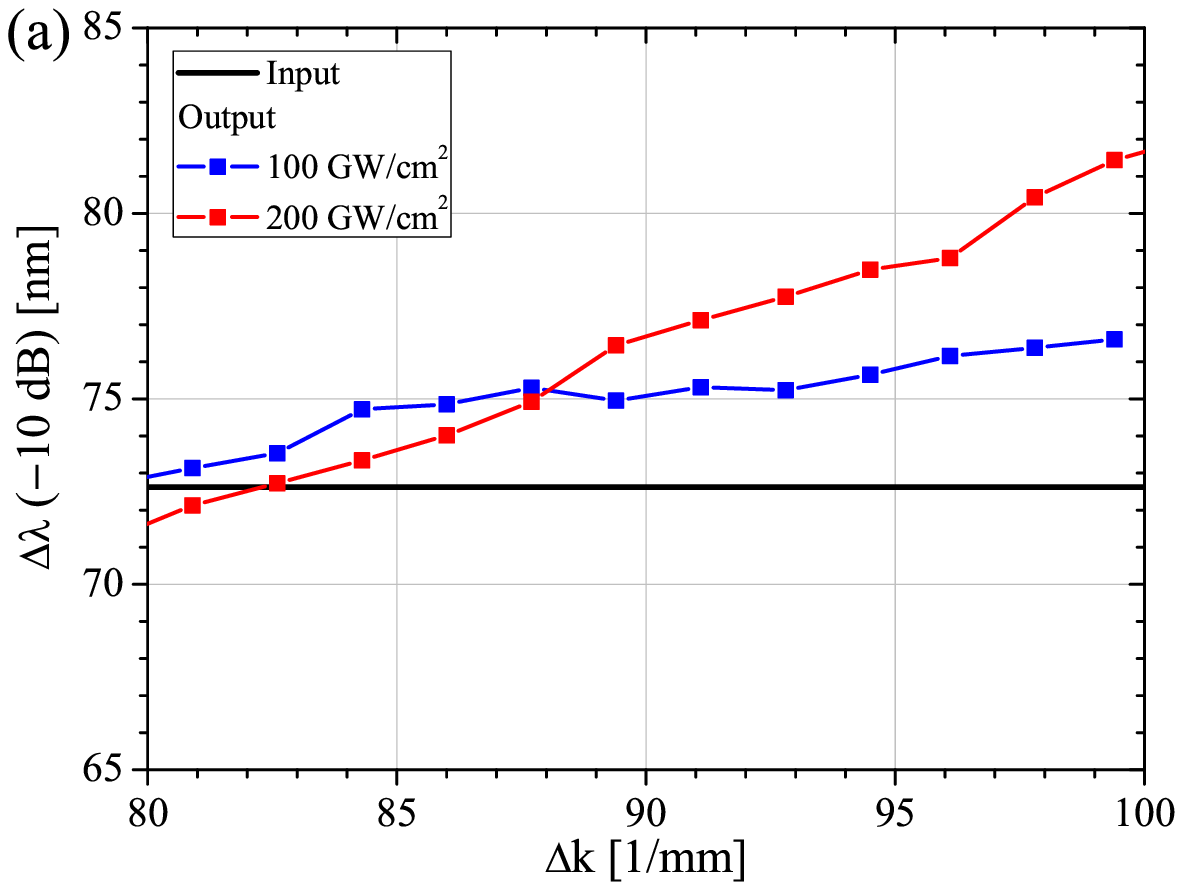}
\includegraphics[width=6cm]{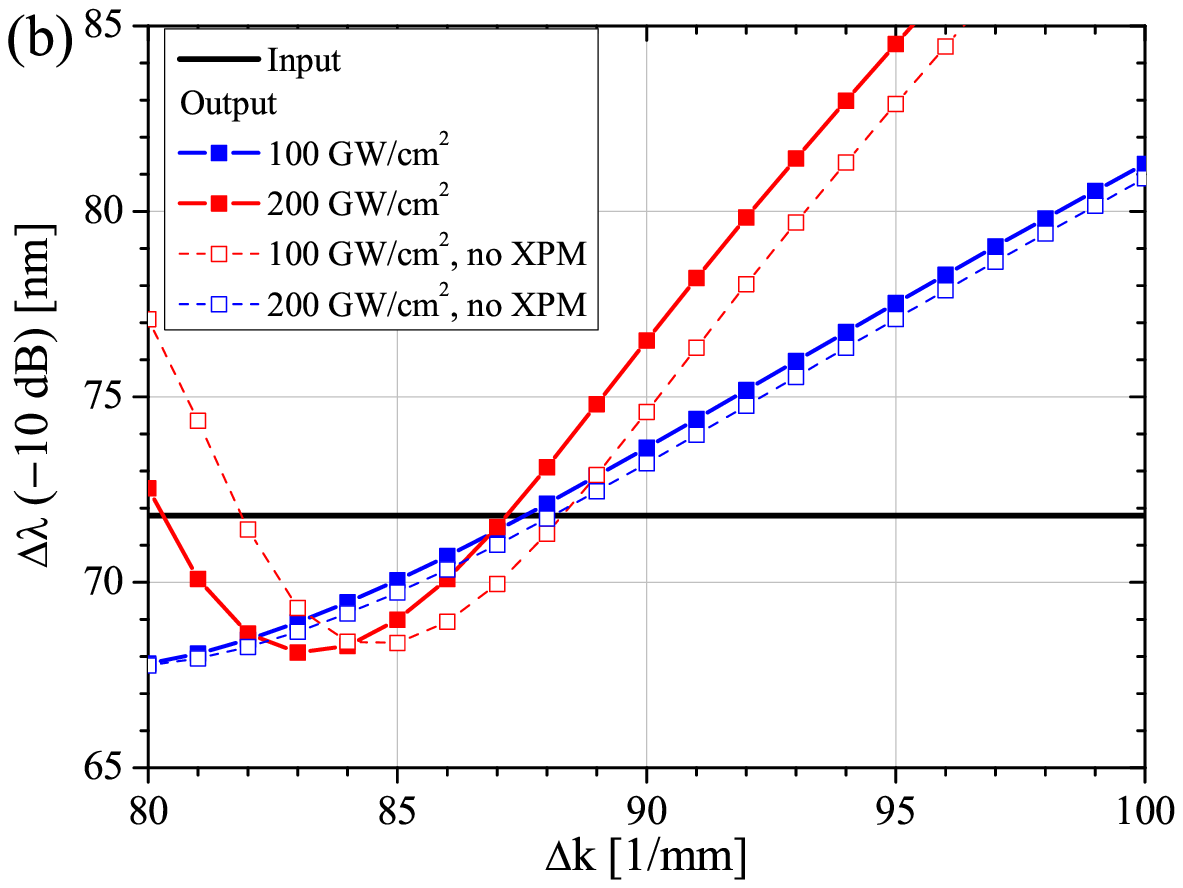}
\includegraphics[width=6cm]{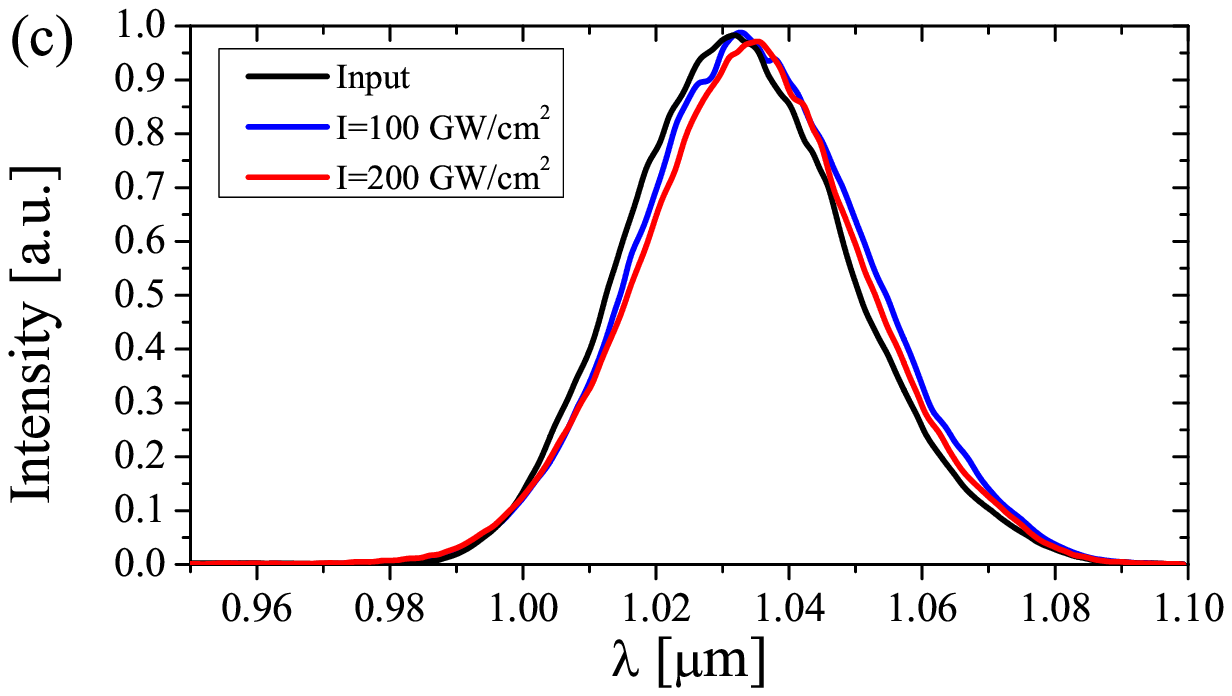}
\includegraphics[width=6cm]{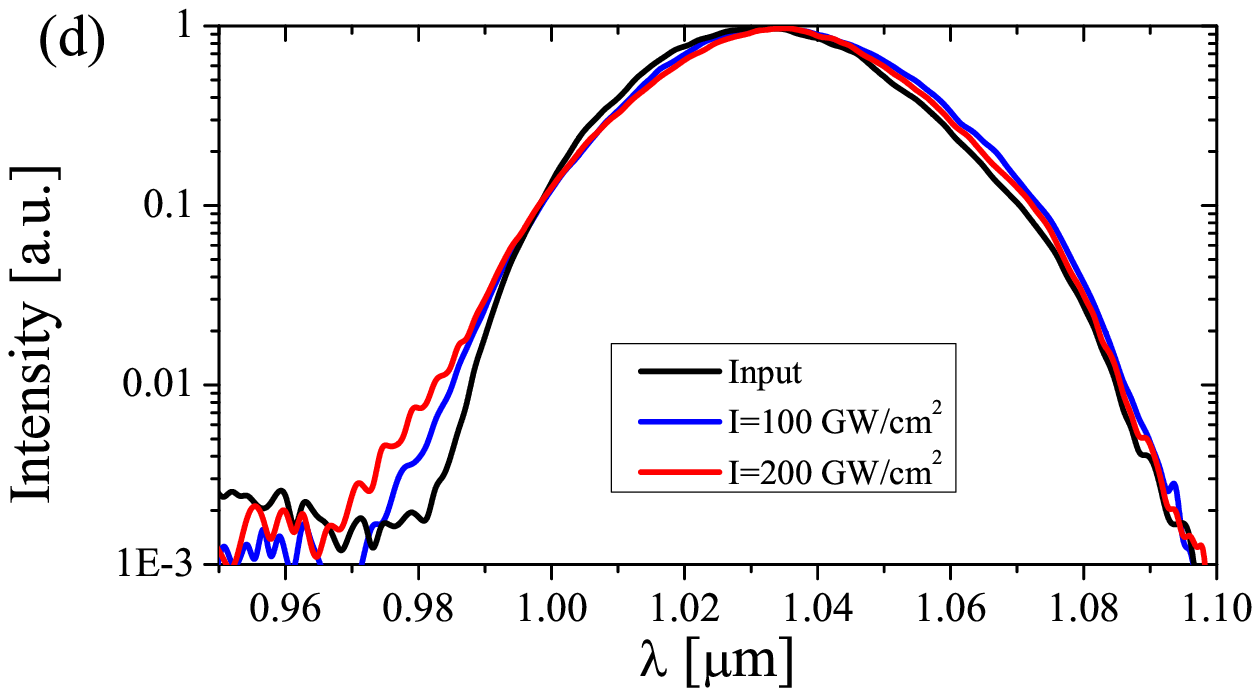}
\includegraphics[width=8cm]{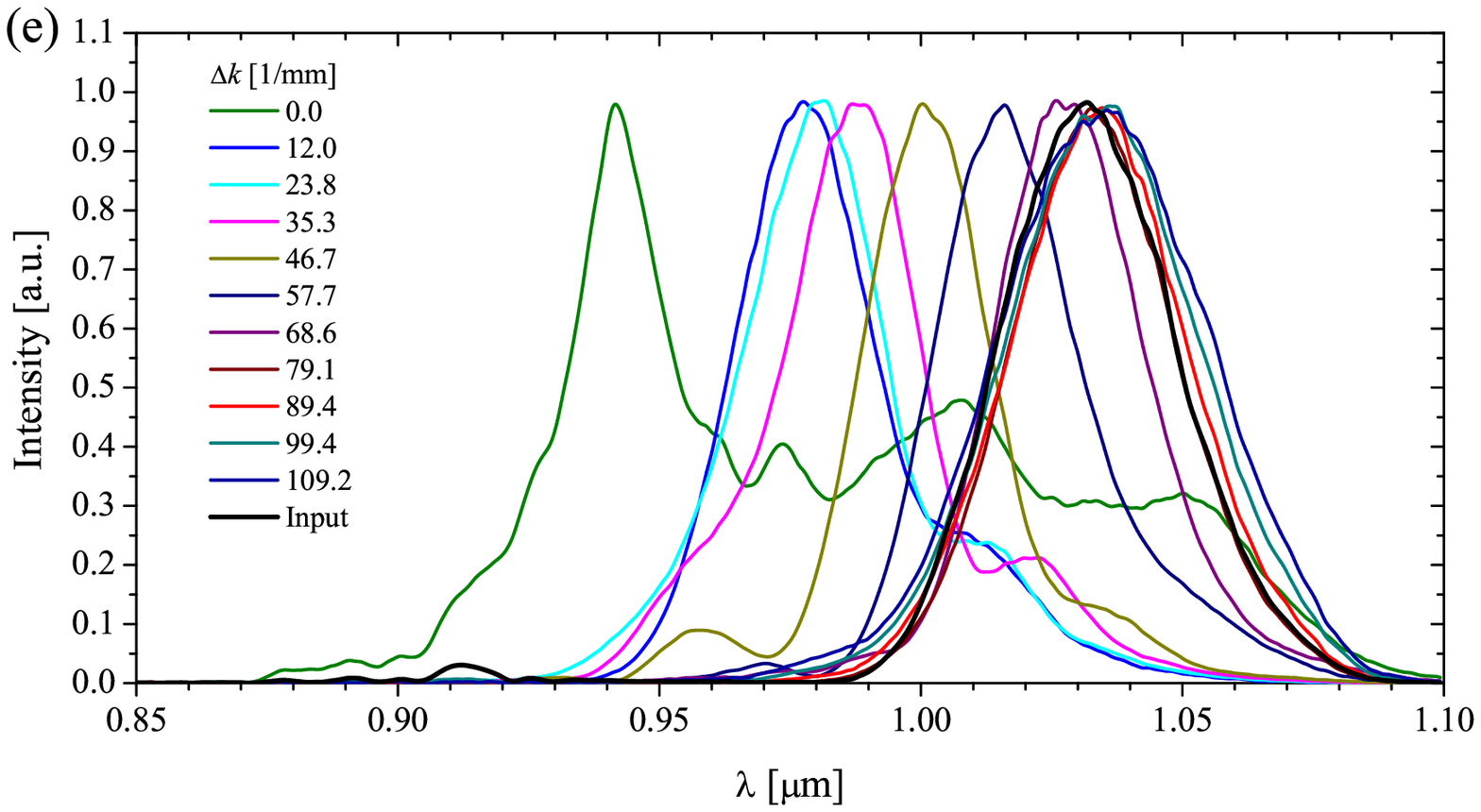}
\end{center}
\caption{\label{fig:BBO-Dkc} (a) Spectral bandwidth at $10$ dB below the peak of the experimentally recorded spectra vs. $\Delta k$ for two different intensities. (b) Similar data from numerical simulations of the coupled SEWA equations \cite{moses:2006,bache:2007} (using the anisotropic quadratic and cubic nonlinear coefficients in Appendix \ref{sec:Crystal-parameters} and Table \ref{tab:summary}, respectively). The input conditions were similar to the experiment (Gaussian pulse with 39.3 nm bandwidth and a group-delay dispersion of $-330~{\rm fs^2}$). (c) and (d) show the spectra on a linear and log scale right at the transition ($\Delta k=88\imm$), where the total SPM is near zero. (e) Various spectra for $I=200~{\rm GW/cm^2}$.}
\end{figure}

In the same spirit, we performed an experiment using the technique employed by Moses et al. The crystal was a 25 mm BBO from Castech with a 10x7 mm$^2$ aperture, cut so $\theta=21^\circ$ and $\phi=-\pi/2$, and antireflection coated. The crystal was pumped with pulses from a 1 kHz Ti:Sapphire regenerative amplifier followed by a commercial OPA, giving $\lambda=1.032 \mic$ pulses with around 40 $\mu$J pulse energy. The pulses were Gaussian with 39.3 nm FWHM bandwidth and an intensity autocorrelation trace revealed a Gaussian pulse with an intensity FWHM of 46 fs; the pulses were therefore nearly transform-limited Gaussian pulses (the time-bandwidth product was 16\% above transform limit). The pump was collimated with a telescope before the crystal and the spot size was measured to 0.5 mm FWHM, and the intensity was controlled by a neutral-density filter. We chose to pump at this wavelength because the GVM-induced self-steepening is smaller than at 800 nm, and besides the transition from negative to positive effective nonlinearity occurs in the so-called stationary regime, where the cascading nonlinearity is non-resonant (see also Sec. \ref{sec:implications} later). We first located the phase-matching angle and then angle-tuned the crystal until any visible nonlinear components of the spectrum had vanished (i.e. it returns close to its input state) as $\ns+\nk\simeq 0$ in this range. By fine-tuning (with $1/6^\circ$ precision) $\theta$ in this range we measured the trends shown in Fig. \ref{fig:BBO-Dkc}(a): this plots the bandwidth 10 dB below the peak value of the recorded spectra for two different intensities, and the curves are seen to cross each other at $\Delta k\simeq 88\imm$. We investigated numerically whether this is a good measure of the transition from negative to positive total nonlinearity; for the parameters used in the numerics the critical point was calculated analytically to be $\Delta k_c=88.5\imm$ (neglecting XPM effects, see later). Fig. \ref{fig:BBO-Dkc}(b) shows the equivalent numerical results, and clearly the two different intensities cross each other very close to the theoretical transition. Two comments are in order: firstly, the model includes XPM effects that give a self-focusing contribution \cite{bache:2007} (see also discussion later in Appendix \ref{sec:NLSE}, where the cascading quintic term is calculated) that for the chosen intensities is quite minimal. We can judge the XPM impact by turning off XPM in the code: the transition now occurs practically at the theoretical value, which is a bit higher than the result with XPM. Thus, we can expect that XPM gives a small correction on the order of $-1\imm$. Secondly, the reason why the spectrum does not return to the input value is that when cascading cancels Kerr SPM effects there are still self-steepening effects that remain (see the detailed experiments by Moses et al. \cite{moses:2007a}, and also the recent discussion by Guo et al. \cite{guo:2012}). Plots (c) and (d) show the spectra right at the transition, and the red edge is quite similar for the two intensities while the blue edge is always higher for the larger intensity. We should mention that we found similar transition "indicators" at very similar $\Delta k$ values by observing the -20 dB bandwidth, FWHM bandwidth, in tracking the wavelength position of the spectral edges, in tracking the center wavelength etc. Especially the latter is the most precise indicator numerically as cascading gives a very pronounced blue-shift of the spectrum -- see graph (e) -- but experimentally the peak was too erratic in the fine-tuning around the transition, which is probably related to the fact that the pump is generated as the SH of the idler of our OPA. We conclude that after correcting for a small XPM contribution the critical transition is found the be $\Delta k_c=89\pm 5\imm$, where the uncertainty lies mostly in finding a suitable transition value to track between the different intensities and phase mismatch values. This critical value corresponds to $\theta=18.5^\circ$, $\rho=2.7^\circ$ and $d_{\rm eff}=2.08$ pm/V, so calculating the cascading nonlinearity and employing the $\ns+\nk=0$ hypothesis, we get
\begin{align}\label{eq:Bache-nKerr11}
    \nk(\omega_1;\omega_1)=4.87\pm 0.44 \cdot 10^{-20} ~{\rm m^2/W}, \; \lambda=1.032\mic
\end{align}
which corresponds to $c_{11}=4.73\pm 0.43\cdot 10^{-22} ~{\rm m^2/V^2}$. The uncertainty might be reduced with a longer pump pulse as self-steepening effects are reduced and it therefore will be easier to compare the nearly-linear spectrum at $\Delta k_c$ (where SPM is nulled and only XPM and self-steepening effects contribute) to the input spectrum, but a drawback in using a longer pulse could lie in the challenge of observing small changes in a narrower spectrum. All in all, we must emphasize that this method is quite accurate for determining the Kerr nonlinearity, especially compared to the 20-30\% uncertainty often seen with other techniques. This is partially because we do not rely on any absolute pump intensity, which can be a large uncertainty, and because the quadratic nonlinearities and the linear dispersion properties of BBO both are very well known.

\begin{figure}[t]
\begin{center}
\includegraphics[width=8.7cm]{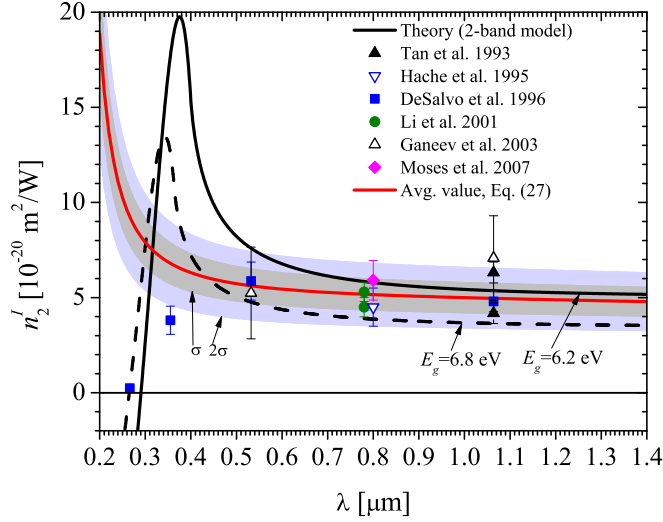}
\end{center}
\caption{\label{fig:nkerr11} Summary of the experimental data for $\nk$-values from the literature corresponding to the $c_{11}$ nonlinear susceptibility coefficient ($\nk=3c_{11}/4n_1^2 \varepsilon_0 c$). The plotted values are the ones reported in Sec. \ref{sec:c11}, i.e. the data values do not necessarily correspond to the ones reported in the literature. References: Tan et al. 1993: \cite{tan:1993}; Hache et al. 1995: \cite{hache:1995}; DeSalvo et al. 1996 \cite{desalvo:1996}; Li et al. 1997 \cite{li:1997}; Li et al. 2001 \cite{Li:2001}; Ganeev et al. 2003 \cite{ganeev:2003}; Moses et al. 2007 \cite{moses:2007a}. The theoretically predicted electronic nonlinearity is calculated with the 2-band model \cite{Sheik-Bahae:1991}. The average value curve was calculated through a weighted mean of the Miller's delta from all data, except the UV measurements below 400 nm, and the shaded areas denoted "$\sigma$" and "$2\sigma$" represent one and two standard deviations, respectively. }
\end{figure}

This completes the $c_{11}$ measurements. A summary of the data is shown in Fig. \ref{fig:nkerr11}, together with the predicted electronic nonlinearity from the two-band model (2BM) \cite{Sheik-Bahae:1991}. The experimental near-IR data are amazingly well predicted on an \textit{absolute} scale by the 2BM; we must here clarify that for the 2BM we chose the material constant $K=3100~\rm eV^{3/2}cm/GW$, which was found appropriate as a single material parameter for dielectrics \cite{desalvo:1996}, and by using the BBO band-gap value $E_g=6.2$ eV \cite{desalvo:1996}. Thus, in practice  there are no free parameters in the model, which underlines the incredible agreement obtained on an absolute scale (and not just the translation from one wavelength to another). We mention here that in DeSalvo et al. \cite{desalvo:1996} they proposed to "rescale" the bandgap do obtain a better fit with the experimental data (mainly because the two-photon absorption values $\beta$ were not accurately predicted with 6.2 eV), and eventually suggested using 6.8 eV instead of 6.2 eV for BBO. This gave a much better $\beta$ agreement at 266 nm, while the 352 nm value was still off. Here we see that the corrected DeSalvo $\nk$ near-IR values actually agree extremely well with the 2BM $\nk$ value when using $E_g=6.2$ eV, while the agreement is less accurate with $E_g=6.8$ eV. A common issue whether one uses 6.2 or 6.8 eV is that the predicted strong enhancement just below 400 nm due to two-photon absorption of the FW is not reflected in the experimental data. This is of little importance for our purpose, because in cascaded SHG the FW wavelength is usually never below 600 nm. 
In order to extract a suitable value for the $c_{11}$ coefficient, we calculated first the Miller's delta from all the data. It is defined as \cite{wynne:1969,Ettoumi:2010}
\begin{align}
    \Delta_{ijkl}&=\frac{\chi^{(3)}_{ijkl}} {\chi^{(1)}_{i}\chi^{(1)}_{j}\chi^{(1)}_{k}\chi^{(1)}_{l}} 
    =\frac{\chi^{(3)}(-\omega_i;\omega_j,\omega_k,\omega_l)} {\chi^{(1)}(\omega_i)\chi^{(1)}(\omega_j)\chi^{(1)}(\omega_k)\chi^{(1)}(\omega_l)}
    \label{eq:Miller}
\end{align}
where $\chi_j^{(1)}(\omega_0)=n_j^2(\omega_0)-1$ is the linear susceptibility. We here excluded the UV measurements below 400 nm as they obviously are not represented well by Miller's scaling. Since Miller's delta should be independent of wavelength this should allow us to calculate an average value based on the measurements at different wavelengths. We confirmed this hypothesis by checking that that Miller's delta vs. measurement wavelength could be fitted to a line with a near-zero slope, before we calculated the weighted average (with a weight given by $\sigma_j^{-2}$ where $\sigma_j$ is the estimated error of the $j$th measurement) as
\begin{align}\label{eq:Delta11_avg}
   \Delta_{1111}=52.3\pm 7.7\times 10^{-24}~{\rm m^2/V^2}
\end{align}
where the standard deviation was calculated using an unbiased weighted average. This value corresponds to $c_{11}=5.0 \cdot 10^{-22} ~{\rm m^2/V^2}$ and $\nk(\omega_1;\omega_1)=5.1\cdot 10^{-20} ~{\rm m^2/W}$ at 800 nm. We see in Fig. \ref{fig:nkerr11} that this average is a quite good representative of the experimental data; all except two are within two standard deviations. We also note that the value we measured in this work fits extremely well with this proposed average value. 

\subsection{Other tensor components}


Banks et al. \cite{Banks:2002} measured the $c_{10}$ tensor component of BBO using third-harmonic generation (THG) measurements at 1053 nm using 350 fs pulses from a 10 Hz Ti:Sapphire regenerative amplifier. The propagation angle was $\theta=37.7^\circ$, and from the data recorded while $\phi$ was varied around zero they extracted $C_{10}=-6\cdot 10^{-24}~\rm m^2/V^2$ using the quadratic nonlinear coefficients of Shoji et al. \cite{Shoji:1999}. (They extracted other values as well, but we choose to use this one because it was calculated from the experimental data using the same effective quadratic nonlinearities of Shoji et al. as we use.) In fact, Banks et al. reported that this value came with a quite large uncertainty due to the uncertainty in the $d_{31}$ -- and $d_{15}$, in absence of Kleinman symmetry -- coefficients in the literature. Since \cite{Banks:2002} define $C_{\mu m}=\chi^{(3)}/4$, we have $C_{\mu m}=c_{\mu m}/4$, so
\begin{align}\label{eq:banks-chi3_10}
    c_{10}({\rm THG})=-0.24\pm0.04\cdot 10^{-22}~\rm m^2/V^2, \; \lambda=1.053\mic
\end{align}
The uncertainty of the measurement was not reported, but for a similar fit it was reported to be $15\%$, which is what we used above. They also measured a three times larger value by using different quadratic nonlinearities, namely $d_{15}=0.16$ pm/V (as opposed to $d_{15}=0.03$ pm/V).

Banks et al. \cite{Banks:2002} also measured the mixture $\tfrac{1}{3}C_{11}\cos^2\theta_m+C_{16}\sin^2\theta_m=4.0\pm0.2 \cdot 10^{-23}~\rm m^2/V^2$ at $\theta_m=47.7^\circ$. 
The TH walk-off angle is $\rho=4.5^\circ$, and replacing in the above expression $\theta\rightarrow \theta+\rho$ will not change significantly the following numbers; therefore we choose to ignore the walk-off contribution because the fit done by Banks et al. also ignored it. If we now use Eq. (\ref{eq:Delta11_avg}), and the apply Miller's scaling to get the $c_{11}$ component appropriate for $\omega+\omega+\omega\rightarrow 3\omega$ interaction, see Eq. (\ref{eq:Miller}), we get $c_{11}(-3\omega;\omega,\omega,\omega)=c_{11}({\rm THG@}\lambda=1.053\mic)=5.24 \cdot 10^{-22}~\rm m^2/V^2$, and therefore 
\begin{align}\label{eq:banksMoses-chi3_16}
    c_{16}({\rm THG})=1.47\pm0.34\cdot 10^{-22}~\rm m^2/V^2, \; \lambda=1.053\mic
\end{align}
A small caveat must be noted: what they measured was related to the $\chi^{(3)}(-3\omega;\omega,\omega,\omega)$ coefficients, because they investigated the yield of the third harmonic with respect to the pump, and decoupled the cascade yield (from multistep SHG mixing, i.e. $\omega+\omega\rightarrow 2\omega$ followed by $\omega+2\omega\rightarrow 3\omega$) to get the pure cubic nonlinear contribution. However, it is not sure that $\chi^{(3)}(-3\omega;\omega,\omega,\omega)= \chi^{(3)}(-\omega;\omega,-\omega,\omega)$, i.e. that the THG nonlinearity is the same as the SPM nonlinearity; this was discussed, e.g., in \cite{bosshard:2000}. Therefore using these values to model the SPM and XPM effects in BBO is an approximation.


Sheik-Bahae and Ebrahimzadeh \cite{sheik-bahae:1997} used the Kerr-lens autocorrelation method \cite{Sheik-Bahae:1997-KLAC} to measure $\nk=3.65\pm 0.6\cdot 10^{-20}~{\rm m^2/W}$ ($\chi^{(3)}=3.43 \pm 0.56\cdot 10^{-22} ~{\rm m^2/V^2}$) using a 76 MHz Ti:Sapphire oscillator giving 120 fs pulses at $\lambda=0.850\mic$. The BBO crystal was cut for phase matching at 800 nm, i.e. with $\theta=29.2^\circ$, and consequently $\rho=3.9^\circ$. The crystal cut was optimized for SHG, so $\phi=-\pi/2$. The pump pulse was $e$-polarized as to ensure lack of phase matching and thereby no cascading. We will now assess whether this assumption is fulfilled. The cascading contributions that might occur are $ee\rightarrow o$ and $ee\rightarrow e$, however the former has zero nonlinearity when $\phi=-\pi/2$, cf. Eq. (\ref{eq:deff-eeo}). The phase mismatch for the latter interaction is $\Delta k_{eee}=396\imm$ for $\theta=29.2^\circ$. The cascading contribution at $\phi=-\pi/2$ then becomes $\cs=-0.51\cdot 10^{-22}~{\rm m^2/V^2}$. 
Thus, the choice of using $e$-polarized light does make the cascading contribution small, but not insignificant (it is roughly equal to the reported uncertainty). If we now correct the measured value with the cascading contributions we get $\chi_{\rm Kerr}^{(3)}=3.94 \pm 0.56\cdot 10^{-22} ~{\rm m^2/V^2}$. 
Let us now understand what tensor components this value represents. By using $e$-polarized light several tensor components come into play for the cubic nonlinearity: the effective Kerr nonlinearity experienced by the pump is an $eeee$ interaction
, which is given by Eq. (\ref{eq:chi3-type0}). For $\theta=29.2^\circ$ the TH walk-off angle is $\rho=4.3^\circ$, and with $\phi=-\pi/2$ the effective nonlinearity is
\begin{equation}\label{eq:bbo-sb1997}
  \chi^{(3)}_{{\rm eff}}(\omega_1;\omega_1)=-1.28 c_{10}+0.48 c_{11}+1.27c_{16}+0.093c_{33}
\end{equation}
Four tensor components appear, and we note that we here have a measurement that involves the $c_{33}$ component. Let us now use the previous results to extract it: using Eq. (\ref{eq:Delta11_avg}), (\ref{eq:banks-chi3_10}) and (\ref{eq:banksMoses-chi3_16}), all suitably converted to 850 nm with Miller's rule, we can from the corrected value of $\chi_{\rm Kerr}^{(3)}=3.94 \pm 0.56\cdot 10^{-22} ~{\rm m^2/V^2}$ and Eq. (\ref{eq:bbo-sb1997}) calculate
\begin{align}\label{eq:BanksMosesSB-nKerr33}
    c_{33}=-5.35\pm8.43 \cdot 10^{-22} ~{\rm m^2/V^2}, \; \lambda=0.85\mic
\end{align}
An exciting consequence of this result is that it points towards a negative, self-defocusing Kerr nonlinearity. However, as indicated the particular value is very uncertain (the uncertainty was calculated with error propagation rules), which mainly stems from the low prefactor in front of the $c_{33}$ term in Eq. (\ref{eq:bbo-sb1997}). Apart from that, it also relies on three other separate measurements,  making it very sensible to the $c_{11}$, $c_{10}$ and $c_{16}$ values used. Here we mention that the large uncertainty of $c_{10}$ reported in \cite{Banks:2002} plays a role; with the various possibilities reported there for the $c_{10}$ value we always get a negative $c_{33}$, and its value can vary by a factor of two from that reported in Eq. (\ref{eq:BanksMosesSB-nKerr33}).

\begin{table}[tb]
  \centering
  \caption{Summary of the literature measurements of the cubic nonlinearities. The column (A) reports the original data and (B) our updated values, if any.}\label{tab:summary}
  \begin{tabular}{c|c|c|c|c|c|c|c|c}
$\lambda$ &	$\theta$ &	$\phi^{(c)}$ &	$\chi_{\rm eff}^{(3)}$ (A) &	 $\chi_{\rm eff}^{(3)}$ (B) &	 $\chi_{\rm eff}^{(3)}$ &	Rep. rate	& $T_{\rm FWHM}$	& Ref. \\
$[\rm nm]$ &	[deg] &	[deg] &	$[{\rm pm^2/V^2}]$ &	$[{\rm pm^2/V^2}]$ & & & & \\				
    \hline
1064 &	22.8 &	($-90$) &	-   &	584$^{(a)}$ &	$c_{11}$ &	? &	30 ps & \cite{tan:1993}	 \\
1064 &	22.8 &	($-90$) & 	420 &	405$^{(a)}$ &	$c_{11}$ &	? &	30 ps & \cite{tan:1993}\\	
800 &	29.2 &	($-90$) &	436 &	-	& $c_{11}$ &	MHz	& 100 fs & \cite{hache:1995}\\	
800 &	26.0 &	($-90$)	&   460 &	541$^{(a)}$ &	$c_{11}$ &	1 kHz &	110 fs	& \cite{moses:2007a} \\	
1064 &	0    &	($-90$) &	270 &	465$^{(b)}$ &	$c_{11}$ &	1 shot &	30 ps & \cite{desalvo:1996}\\	
532 &	0    &	($-90$) &	522 &	582$^{(b)}$ &	$c_{11}$ &	1 shot &	30 ps & \cite{desalvo:1996}\\	
355 &	0    &	($-90$) &	354 &	392$^{(b)}$ &	$c_{11}$ &	1 shot &	30 ps & \cite{desalvo:1996}\\	
266 &	0    &	($-90$) &	26 &	- &	$c_{11}$ &	1 shot &	30 ps & \cite{desalvo:1996}\\	
532 &	0    &	0       &	476 &	537$^{(b)}$ &	$c_{11}$ &	10 Hz &	25 ps  & \cite{li:1997} \\	
532 &	0    &	90      &	397 &	458$^{(b)}$ &	$c_{11}$ &	10 Hz &	25 ps	& \cite{li:1997} \\	
1064 &	0    &	?       &	484 &	679$^{(b)}$ &	$c_{11}$ &	10 Hz &	25 ps  & \cite{li:1997} \\	
780 &	0    &	0       &	391 &	518$^{(b)}$ &	$c_{11}$ &	76 MHz$^{(d)}$ &	 150 fs  & \cite{Li:2001} \\	
780 &	0    & 	90      &	312 &	439$^{(b)}$ &	$c_{11}$ &	76 MHz$^{(d)}$ &	 150 fs	& \cite{Li:2001} \\	
1064 &	51.0 &	($-90$) &	717 &	692$^{(b)}$ &	$c_{11}$ &	2 Hz &	55 ps	 & \cite{ganeev:2003}\\	
532 &	51.0 &	($-90$) &	794 &	573$^{(b)}$ &	$c_{11}$ &	2 Hz &	55 ps	 & \cite{ganeev:2003}\\	
1032 &	21.0 &	$-90$	&   473 &	-           &	$c_{11}$ &	1 kHz &	46 fs	& This work \\	
1053 &	37.7 &	0       &	-72$^{(e,f)}$ &	-   &	$c_{10}$ &	10 Hz &	350 fs 	& \cite{Banks:2002}\\	
1053 &	37.7 &	0       &	-24$^{(g,f)}$ &	-   &	$c_{10}$ &	10 Hz &	350 fs 	& \cite{Banks:2002}\\	
1053 &	47.7 &	0       &	160$^{(f)}$ &	-           &	$c_{11}$, $c_{16}$$^{(h)}$  &	 10 Hz &	350 fs	 & \cite{Banks:2002}\\
850 &	29.2 &	($-90$) &	343 &	394$^{(b)}$ &	Eq. (\ref{eq:chi3-type0}) &	 76 MHz &	 120 fs	& \cite{sheik-bahae:1997}\\
    \hline
  \end{tabular}
  {\footnotesize $^{(a)}$The corrected data used updated nonlinearities. $^{(b)}$The corrected data adjusted for cascading contributions. $^{(c)}$The parenthesis indicates that the angle was not reported, so the value shown was the angle we believe was used. $^{(d)}$The repetition rate was lowered to 760 KHz with the same result. $^{(e)}$Fit using quadratic nonlinearities of \cite{fan:1989}. $^{(f)}$The THG tensor component was measured instead of the cubic self-action components. $^{(g)}$Fit using quadratic nonlinearities of \cite{Shoji:1999}. $^{(h)}$The measured mixture was $\tfrac{1}{3}c_{11} \cos^2\theta+c_{16}\sin^2\theta$.}
\end{table}

\subsection{Summary of experiments}
\label{sec:cij_summary}

A summary of the data reported in the literature along with our corrected or updated values is shown in Table \ref{tab:summary}. This table also gives an overview of the crystal angles, pulse duration, repetition rates, wavelengths and the tensor components accessed in the measurements.

\begin{table}[tb]
  \centering
  \caption{
  Proposed nonlinear susceptibilities for the BBO anisotropic Kerr nonlinearity. Note the $c_{16}$ value is deduced from the $c_{11}$ and $c_{10}$ coefficients, and the $c_{33}$ value is deduced from the $c_{11}$, $c_{10}$ and $c_{16}$ coefficients. The Miller's delta $\Delta_{ijkl}$ are calculated from Eq. (\ref{eq:Miller}).}\label{tab:cij}
  \begin{tabular}{c|c|c|c|l}
 &	$\lambda$ & $\chi^{(3)}$ &	$\Delta_{ijkl}$ &	Ref. \\
 & $[\mu \rm m]$ & $[10^{-24}~\rm m^2/V^2]$ & $[10^{-24}~\rm m^2/V^2]$ & \\
    \hline
$c_{11}=\chi^{(3)}_{XXXX}$ &	- &   - &	$52.3\pm 7.7$$^{(a)}$	& Eq. (\ref{eq:Delta11_avg}) \\
$c_{10}=\chi^{(3)}_{XXYZ}$ &	1.053 &	$-24\pm 4$&	 -3.04 & \cite{Banks:2002}$^{(b)}$	\\
$c_{16}=\chi^{(3)}_{XXZZ}$ &	1.053 &	$147\pm 34$ &	23.8	& \cite{Banks:2002}$^{(b)}$\\
$c_{33}=\chi^{(3)}_{ZZZZ}$ &	0.850 &	$-535\pm843$ &	 -147 	& \cite{sheik-bahae:1997}\\
    \hline
  \end{tabular}
  {\\\footnotesize $^{(a)}$This value corresponds to $c_{11}=5.0 \cdot 10^{-22} ~{\rm m^2/V^2}$ at 800 nm.  $^{(b)}$The THG tensor component was measured instead of the cubic self-action components.}
\end{table}

Our analysis of the experiments \cite{hache:1995,desalvo:1996,sheik-bahae:1997,Li:2001,Banks:2002,ganeev:2003,moses:2007a} points towards using the Kerr nonlinearities summarized in Table \ref{tab:cij}. Only the susceptibilities are reported in order to minimize the errors when using the values for composite nonlinear indices. We have there also indicated the Miller's delta calculated from Eq. (\ref{eq:Miller}).  
Since the experiments are performed at different wavelengths, using the $\Delta_{ijkl}$ values makes it easier to evaluate a linear combination of the nonlinear coefficient at some particular wavelength. [Another popular model for scaling the cubic nonlinearity is the Boling-Glass-Owyoung (BGO) model \cite{Boling:1978}, but we did not see a big difference in using that model compared to Miller's rule. Besides, extending the BGO model to anisotropic nonlinearities is not straightforward, so we prefer to use Miller's rule for frequency scaling.] The values are reported with more significant digits than supported by the uncertainties, but this is done on purpose so it is easier to cross-check the coefficients as well as frequency scaling them.

\begin{figure}[t]
\begin{center}
\includegraphics[width=8.7cm]{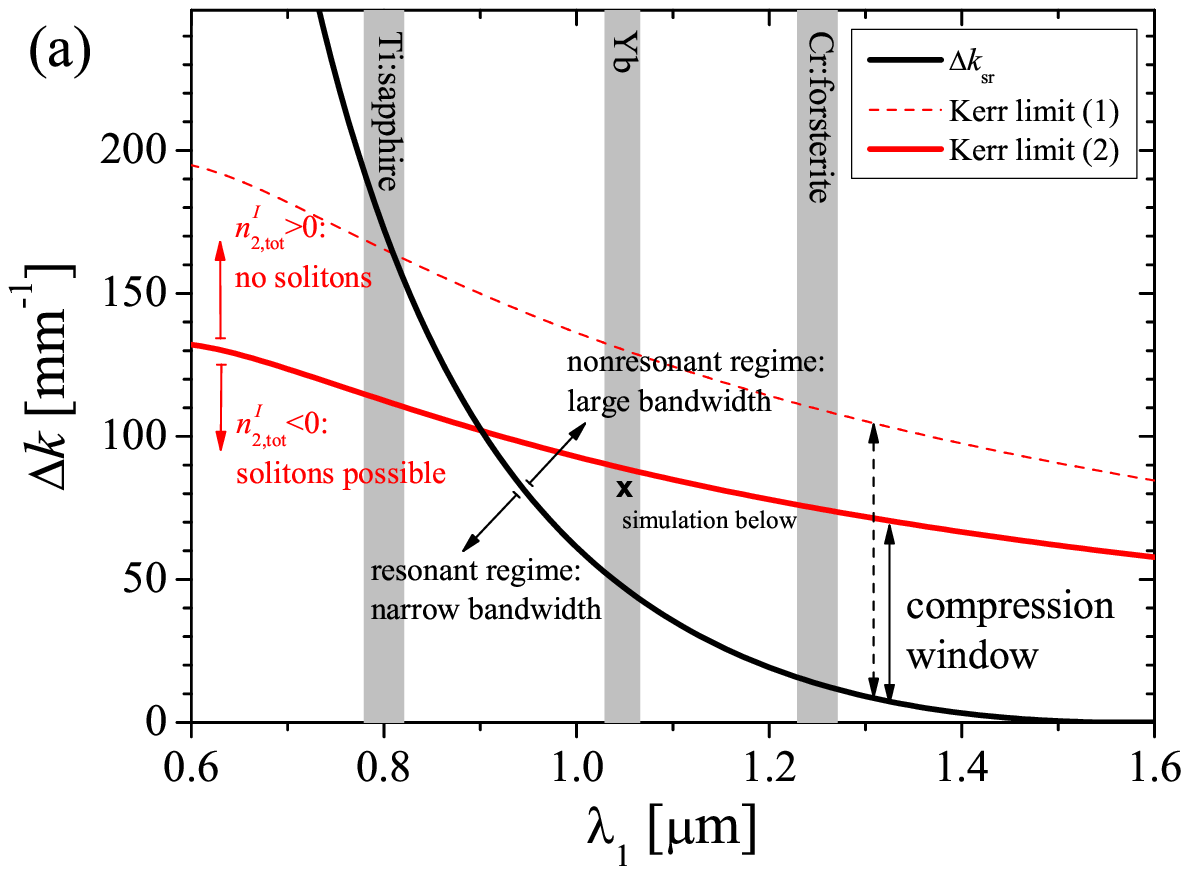}
\includegraphics[width=6cm]{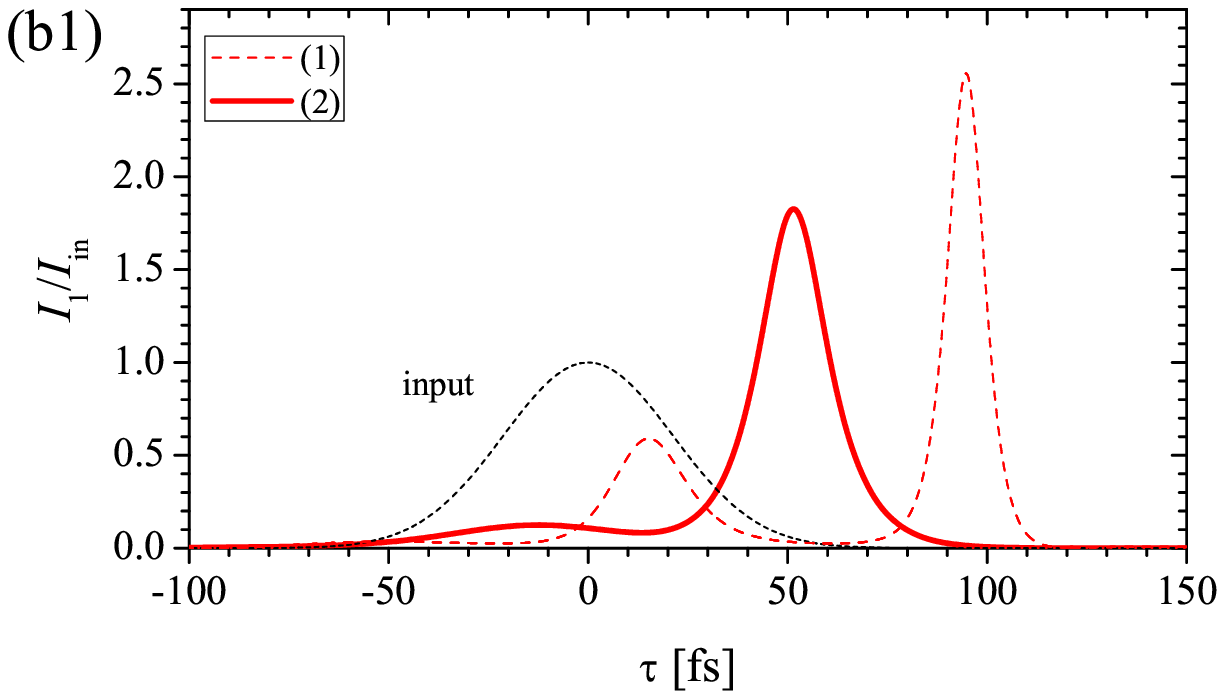}
\includegraphics[width=6cm]{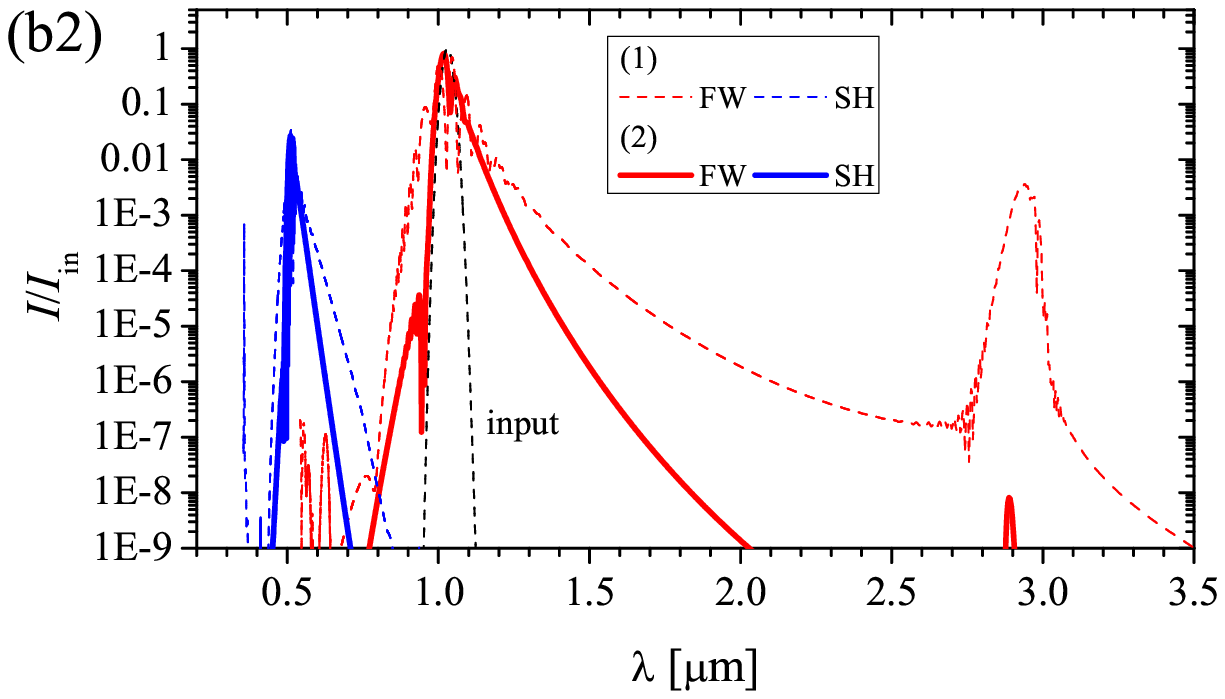}
\end{center}
\caption{\label{fig:cw} (a) Compression diagram for BBO type I cascaded SHG. 
In order to excite solitons $\Delta k$ must be kept below the Kerr limit (red line). Optimal compression occurs when the cascaded nonlinearities dominate over GVM effects ($\Delta k>\Delta k_{\rm sr}$, above the black line).  Note that $\Delta k_{\rm sr}$ is calculated for the full dispersion case, and that for $\lambda_1>1.49~\mu$m the FW GVD becomes anomalous.  
We have also indicated the operation wavelengths of Cr:forsterite, Yb and Ti:Sapphire based amplifiers. The Kerr limit employs Miller's rule to calculate the nonlinear quadratic and cubic susceptibilities at other wavelengths, case (1) corresponds to the 'old' Kerr value $\nk(\omega_1;\omega_1)=3.65 \cdot 10^{-20}~{\rm m^2/W}@850$ nm (taken from \cite{sheik-bahae:1997}) and case (2) corresponds to the Kerr value proposed in this work $\Delta_{1111}=52.3~{\rm m^2/V^2}$, corresponding to $\nk(\omega_1;\omega_1)=5.06 \cdot 10^{-20}~{\rm m^2/W}@850$ nm. (b) Numerical simulation of the case marked with '\textsf{x}' in (a): a 50 fs@1030 nm $I_{\rm in}=500~{\rm GW/cm^2}$ pulse propagating in a 25 mm BBO crystal with $\theta=19.1^\circ$, $\rho=2.8^\circ$ and $\phi=-90^\circ$ ($\Delta k=80~{\rm mm^{-1}}$).  The simulations used the plane-wave SVEA equations (\ref{eq:FW-int})-(\ref{eq:SH-int}) including full dispersion and extended to include self-steepening, and case (1) assumes an isotropic Kerr nonlinearity and $\nk(\omega_1;\omega_1)=3.65 \cdot 10^{-20}~{\rm m^2/W}@850$ nm, while (2) uses the anisotropic coefficients of Table \ref{tab:cij}, and Eq. (\ref{eq:Miller}) to calculate the nonlinear coefficients at 1030 nm.}
\end{figure}

\subsection{Implications for cascaded pulse compression in BBO}
\label{sec:implications}

In context of cascaded quadratic nonlinearities by far most important component is the FW SPM coefficient $\chi^{(3)}_{\rm  eff}(\omega_1;\omega_1)$, and in a type I $oo\rightarrow e$ configuration it is given by the $c_{11}$ component. 
In some of our previous work related to BBO \cite{bache:2007a,bache:2007,bache:2008}, we assumed an isotropic Kerr nonlinearity, and have used the value $\nk(\omega_1;\omega_1)=3.65\cdot 10^{-20}~{\rm m^2/W}$ taken from \cite{sheik-bahae:1997}, and in a later publication, we have used the value given by Eq. (\ref{eq:Moses-nKerr11}). An implication of this larger value is that the "compression window" \cite{bache:2007} becomes smaller. With this we imply a range of phase-mismatch values, where compression is optimal. We first need $\Delta k$ to be small enough so $n_{2,\rm tot}^{I}<0$, i.e. $|\ns|>\nk(\omega_1;\omega_1)$. This is the upper end of the window, also denoted the Kerr limit. Note that as mentioned in Sec. \ref{sec:anisotropy} the Kerr limit is quite sensitive to whether one includes the spatial walk-off angle $\rho$ in calculating the $d_{\rm eff}$-value in the cascading expression. Once below the Kerr limit, decreasing $\Delta k$ strengthens the total defocusing nonlinearity. However, at a certain point the group-velocity mismatch (GVM) becomes too strong and the compression quality is strongly reduced. Several things happen: (a) the GVM-induced self-steepening term \cite{moses:2006,guo:2012} is increased as it scales as $d_{12}/\Delta k$, where $d_{12}$ is the GVM parameter, so the compressed pulse experiences a strong pulse-front shock. (b) When $\Delta k$ becomes too low (specified more accurately below) the SH will experience a resonant phase matching condition of a sideband frequency to the center frequency of the FW spectrum \cite{zeng:2012}. This is damaging to pulse compression because in the cascading process effectively this is the cascading bandwidth felt by the FW, and in the transition from the nonresonant to the resonant regime, the bandwidth essentially goes from being octave spanning to becoming resonant and thereby narrow \cite{zhou:2012}. The criterion for being in the nonresonant regime, also denoted the stationary regime, can in the simple case where only second-order dispersion is considered \cite{bache:2007a} be expressed as $d_{12}^2-2k_2^{(2)}(\omega_2)\Delta k<0$, where $k_2^{(2)}(\omega_2)$ is the group-velocity dispersion (GVD) coefficient of the SH. Thus, in the case where the SH GVD is normal (positive) we have the stationary (nonresonant) regime with broadband cascading when
    $\Delta k>\Delta k_{\rm sr}\equiv d_{12}^2/[2k_2^{(2)}(\omega_2)]$.
Below this threshold, in the so-called nonstationary regime, the cascaded nonlinearity is as mentioned resonant: the poor bandwidth implies that there is no possibility to achieve few-cycle duration \cite{bache:2010}. At the same time the compressed pulse quality is low as there is a strong pulse shock front: this stems from the $d_{12}/\Delta k$ ratio being large, and this term controls the cascading-induced self-steepening, as mentioned above. As a consequence one can only achieve a decent pulse quality by keeping very low soliton orders (roughly below 2) \cite{bache:2010}. The implications of the larger nonlinear SPM Kerr nonlinearity for the FW is visualized in Fig. \ref{fig:cw}(a). With the lower value, $\nk(\omega_1;\omega_1)=3.65\cdot 10^{-20}~{\rm m^2/W}@ 850$ nm, a large compression window is predicted, almost even encompassing Ti:Sapphire laser wavelengths. With the new larger value, $\Delta_{1111}=52.3~{\rm m^2/V^2}$ corresponding to $\nk(\omega_1;\omega_1)=5.06\cdot 10^{-20}~{\rm m^2/W}@850$ nm, the compression window starts opening around 900 nm. 
To judge the impact of using this larger Kerr value, Fig. \ref{fig:cw}(b) shows a simulation at 1030 nm, where the phase mismatch is chosen to lie inside the compression window using the new larger Kerr value. After 25 mm of propagation the short (50 fs) and intense ($I_{\rm in}=500~{\rm GW/cm^2}$) input pulse is compressed somewhat due to the formation of a self-defocusing soliton, and the spectrum shows SPM-like broadening and even formation of a soliton-induced Cherenkov wave (dispersive wave) around 2900 nm \cite{bache:2010e}. When instead using the 'old' lower Kerr value (dashed lines) the result is changed quite a lot: this is because $|n_{2,\rm tot}|$ is now much larger so the intensity gives a larger effective soliton order $N_{\rm eff}^2\propto |n_{2,\rm tot}|I_{\rm in}$ (the specific numbers are $N_{\rm eff}=3.4$ in case (1) and $N_{\rm eff}=1.7$ in case (2). The soliton is therefore compressed more and earlier in case (1), so after 25 mm soliton splitting has already occurred and in the spectrum the broadening is more pronounced and the dispersive wave is much stronger. These results clearly show how sensible cascading is to the value of the FW Kerr SPM parameter.

The experiments \cite{liu:1999,ashihara:2002} carried out at 800 nm were performed well into the nonstationary region, which was necessary to achieve a defocusing nonlinearity. No few-cycle compressed solitons were observed in \cite{ashihara:2002}, as the intensity had to be kept low in order to avoid severe GVM-induced self-steepening. (Note that the simulations in \cite{ashihara:2002} used $\chi^{(3)}=5\cdot 10^{-22}~{\rm m^2/V^2}@800$ nm, i.e. identical to the value we suggest.) Instead the experiment carried out at 1250 nm \cite{moses:2006b} was done in the stationary regime, and indeed a few-cycle soliton was observed. 

A question is whether pumping with $e$-polarized light could give any advantages. As the $ee\rightarrow o$ interaction is heavily phase mismatched in a negative uniaxial crystal, the main cascading channel would be the $ee\rightarrow e$ interaction. As expected this interaction can never be phase matched. The maximum $d_{\rm eff}$ is for $\theta=0$, and this turns out to give the maximum cascading strength as well. However, our calculations show that it is too small to overcome the $\nk(\omega_1;\omega_1)$ nonlinearity, and the total effective cubic nonlinearity would be focusing. If $\theta=\pi/2$ we get $d_{\rm eff}=d_{33}$, which for BBO is very small (this is typical for borates, while the niobates instead have very large $d_{33}$ components). Nevertheless, for $\theta=\pi/2$ the $c_{33}$ component determines the SPM coefficient, and as we found it could be negative. The total effective cubic nonlinearity would therefore be de-focusing. Investigating a BBO crystal pumped with $\theta=\pi/2$ using $e$-polarized light would therefore be an interesting next step to resolve this issue.

\begin{figure}[t]
\includegraphics[height=4cm]{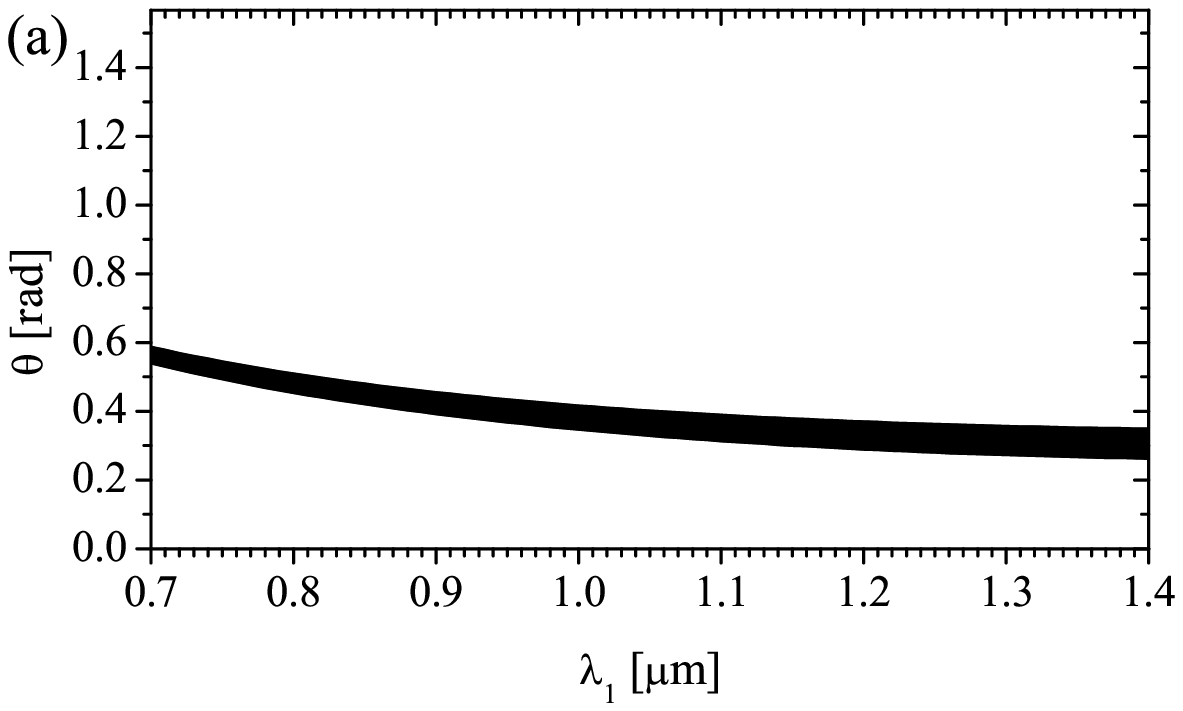}
\includegraphics[height=4cm]{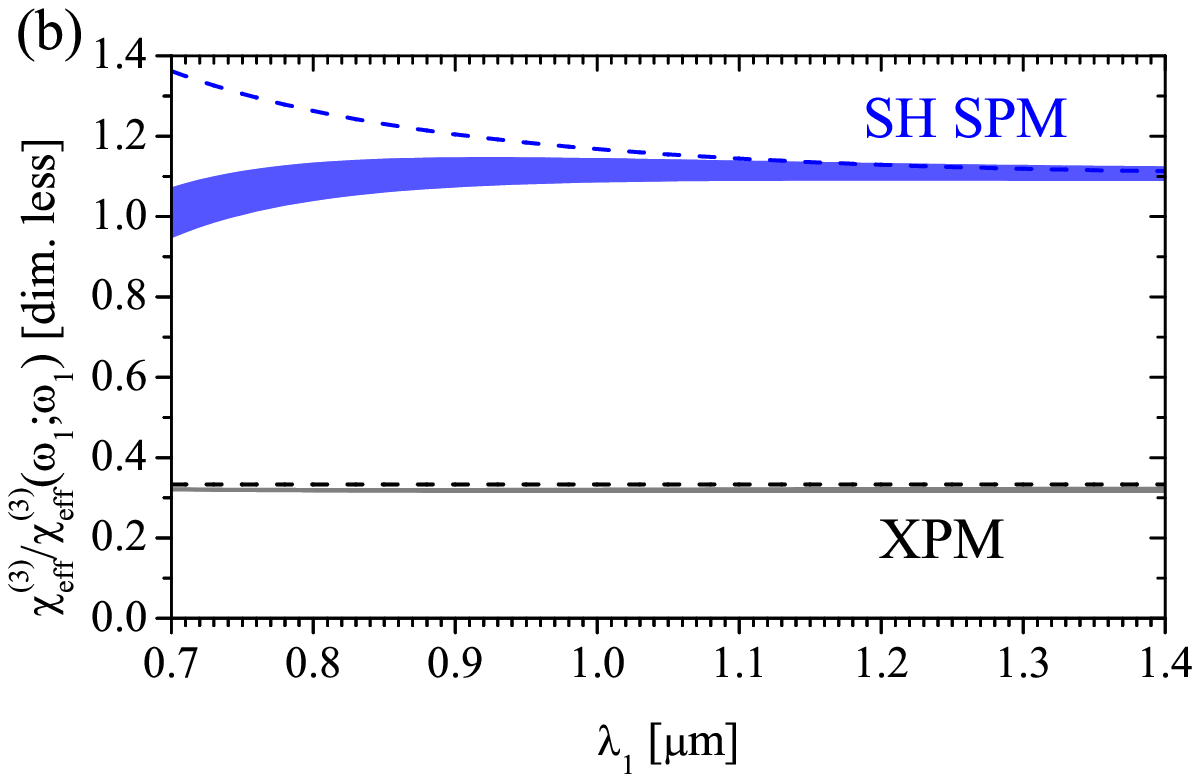}
\caption{\label{fig:spm-xpm} Estimating the operating parameters vs. FW wavelength for cascaded SHG in BBO. (a) The $\theta$-range for which $\Delta k>0$ is achieved as well as $|\ns|>\nk(\omega_1;\omega_1)$. (b) The XPM term $\chi_{\rm eff}^{(3)}(\omega_1;\omega_2)$ and SH SPM term $\chi_{\rm eff}^{(3)}(\omega_2;\omega_2)$, both normalized to the FW SPM term $\chi_{\rm eff}^{(3)}(\omega_1;\omega_1)$. The range indicated corresponds to the $\theta$ range in (a). The dashed lines indicate the isotropic limit. All nonlinear coefficients are scaled to other wavelengths using Miller's rule, and we used the SHG nonlinearities at 1064 nm from App. \ref{sec:Crystal-parameters} as well as the cubic nonlinear parameters listed in Sec. \ref{sec:summary}. We also took $\phi=-\pi/2$.}
\end{figure}

We should finally mention that the error made in assuming an isotropic response for the cascaded soliton compression is not large: most importantly, the crucial parameter is the FW SPM coefficient, which as we saw is the same in the isotropic and in the anisotropic cases for this type I interaction. The XPM and SH SPM coefficients change when anisotropy is taken into account, but they play a minor role and rather tend to perturb the result more than shape it. In order to make a more quantitative statement, we calculated the $\theta$-ranges where self-defocusing solitons can be excited in BBO. A necessary (but not sufficient \cite{bache:2007}) criterion for this to happen is $\Delta k>0$ as well as $\Delta k$ low enough for $|\ns|>\nk(\omega_1;\omega_1)$. The range is shown vs. FW wavelength in Fig. \ref{fig:spm-xpm}(a). Inside this $\theta$ range we can now for each wavelength calculate the anisotropic Kerr nonlinearities. As we know the FW SPM coefficient does not change with $\theta$ (but it does change across the wavelength range shown, which we here estimate using Miller's rule), so we can normalize the results to this value. We then get the results shown in Fig. \ref{fig:spm-xpm}(b). The XPM term is seen to lie close to the isotropic value $1/3$ (found by taking $\chi_{\rm eff}^{(3)}(\omega_1;\omega_2)=c_{11}/3$, dashed grey line). Instead the SH SPM term lies well below the isotropic value [found taking $\chi_{\rm eff}^{(3)}(\omega_2;\omega_2)=c_{11}$, dashed blue line]. The main reason for this is the large negative $c_{33}$ component. In summary, assuming an isotropic response in BBO is a good approximation for the FW XPM term, mainly because the working $\theta$ range lies close to zero (the isotropic limit). This result cannot be generalized to other nonlinear crystals, where the working range might lie closer to $\pi/2$; an example is lithium niobate that operates quite close to $\theta=\pi/2$ in the type I configuration, see e.g. \cite{bache:2010}. For the SH SPM term its value is quite far from the isotropic case, but off all the parameters this is the least important one because the SH SPM is negligible in the cascading limit compared to FW XPM and in particular FW SPM. To confirm this we checked that the simulations of case (2) in Fig. \ref{fig:cw}(b) were almost identical results when using isotropic Kerr nonlinearities. A similar conclusion was drawn in our recent work, where the anisotropy was also taken into account \cite{guo:2012}.

\section{Summary}\label{sec:summary}

We have analyzed the Kerr nonlinear index in BBO from a number of experiments \cite{hache:1995,desalvo:1996,sheik-bahae:1997,Li:2001,Banks:2002,ganeev:2003,moses:2007a}, and we argued that in many of them contributions from cascaded SHG nonlinearities need to be taken into account. We also performed a very accurate measurement of the main tensor component $c_{11}$ by balancing the Kerr nonlinearity with a negative cascading nonlinearity. Our analysis points towards using the Kerr nonlinearities summarized in Table \ref{tab:cij}. They encompass all four relevant tensor components for describing the anisotropic nature of the BBO Kerr nonlinearity under Kleinman symmetry. The most reliable value is the $c_{11}$ component, relevant for describing the $oooo$ SPM coefficient, while the other three are more uncertain: the $c_{16}$ and $c_{10}$ values are measured for THG, so using them for describing SPM and XPM effects is an approximation. Finally, the $c_{33}$ coefficient is very uncertain as it was measured in a non-ideal way (since the experiment used a crystal angle where its importance is very small) and on top of that all the three other coefficients in the table were used to deduce its value. In all cases the measured coefficients had contributions from both instantaneous electronic and delayed Raman effects, but we assumed that the Raman contributions to the measured nonlinearities were vanishing.

The proposed $c_{11}$ value was calculated as a weighted average over 14 different experiments in the 532-1064 nm range, and is therefore expressed as a wavelength-independent Miller's delta $\Delta_{1111}=52.3\pm 7.7\times 10^{-24}~{\rm m^2/V^2}$, cf. Eq. (\ref{eq:Delta11_avg}); at 800 nm this value corresponds to $c_{11}=5.0\times10^{-22}~\rm m^2/V^2$ and $\nk=5.1\cdot 10^{-20} ~{\rm m^2/W}$, which is larger than typical values used in the literature. We stress that the $c_{11}$ value we measured in this work agrees extremely well with this proposed value, and we believe that this method -- where the focusing Kerr nonlinearity is nulled by an equivalent but defocusing cascading nonlinearity -- is quite precise and can be used in many other contexts as an alternative to Z-scan measurements. 
When plotting the (corrected) Kerr nonlinear refractive indices vs. wavelength, see Fig. \ref{fig:nkerr11}, we came to a surprisingly good \textit{absolute }agreement with the 2-band model \cite{Sheik-Bahae:1991}, except in the short end of the visible range. 

Finally we showed the predicted consequences for cascaded femtosecond pulse compression exploiting type I interaction in BBO when using this larger value: the operation range where few-cycle pulse compression can be achieved will shift to longer wavelengths, roughly above $1.0\mic$. We also showed that the anisotropic XPM coefficient of the FW is quite similar to the isotropic one, but this relies on the particular range of phase-mismatch rotation angles used in BBO, so this result cannot be generalized to other crystals exploiting type I interaction. Instead the anisotropic SH SPM coefficient is quite different from the one found assuming an isotropic Kerr nonlinearity, but since in cascading the SH is weak this difference will amount to much in a simulation of cascaded SHG.

\section{Acknowledgments}

\label{sec:Acknowledgements}

The Danish Council for Independent Research (grants no. 21-04-0506, 274-08-0479, and 11-106702) is acknowledged for support.  Xianglong Zeng acknowledges the support of Marie Curie International Incoming Fellowship from EU (grant No. PIIF-GA-2009--253289) and the financial support from National Natural Science Foundation of China (60978004) and Shanghai Shuguang Program (10SG38). Frank W. Wise, Jeffrey Moses and Satoshi Ashihara are acknowledged for useful discussions. We would also like to thank some of the authors behind the measurements of the Kerr nonlinearity in the literature for comments and additional details: Mansoor Sheik-Bahae, Fran\c{c}ois Hache, Ji Wei, Tan Huiming, and Rashid Ganeev.

\appendix

\section{The propagation equations under the slowly-varying envelope approximation}
\label{sec:SVEA}

We have previously touched upon the issue of anisotropic cubic nonlinearities in quadratic nonlinear crystals \cite{bache:2010}, where we focused on type I interaction and the measurements in the literature for the lithium niobate crystal. Here we formulate the complete propagation equations that hold for both type 0 and type I, and discuss the anisotropic cubic tensor components relevant for the BBO crystal class.

Let us show briefly how to derive the basic slowly-varying envelope equations (SVEA) for degenerate SHG (i.e. where the FW photons are degenerate), thus describing both type 0 and type I SHG. We do this in order to avoid confusion about how to define the relation between the electric field and the susceptibility tensors. Working in mks units, from Maxwell's equations and applying the paraxial approximation we immediately get the time-domain wave equation
\begin{align}\label{eq:we}
\nabla^2 \mathbf{E}(\mathbf{r},t)=\mu_0 \frac{\partial^2}{\partial t^2}\mathbf{D}(\mathbf{r},t)
\end{align}
where $\mathbf{D}=\varepsilon_0 \mathbf{E}+ \mathbf{P}$ is the displacement field, and $\mathbf{P}$ is the induced polarization. 
In frequency domain, where $\mathbf{E}(\mathbf{r},\omega)=\int_{-\infty}^\infty {\rm d}t e^{i \omega t}\mathbf{E}(\mathbf{r},t)$ and similarly for $\mathbf{P}(\omega)$, we may separate the linear and nonlinear polarization response as
\begin{align}\label{eq:P-separation}
\mathbf{P}(\mathbf{r},\omega)=\varepsilon_0\underline{\chi}^{(1)}(\omega)\cdot \mathbf{E}(\mathbf{r},\omega)+\mathbf{P}_{\rm NL}(\mathbf{r},\omega),
\end{align}
where $\underline{\chi}^{(1)}(\omega)$ is the linear susceptibility tensor. The nonlinear part is usually expanded in a power series as
\begin{align}\label{eq:PNL-expansion}
\varepsilon_0^{-1}\mathbf{P}_{\rm NL}(\mathbf{r},\omega)= \underline{\chi}^{(2)}(\omega):\mathbf{E}(\mathbf{r},\omega)\mathbf{E}(\mathbf{r},\omega)+
\underline{\chi}^{(3)}(\omega)\vdots \mathbf{E}(\mathbf{r},\omega)\mathbf{E}(\mathbf{r},\omega)\mathbf{E}(\mathbf{r},\omega)+\cdots
\end{align}
The tensor products are calculated by summation over the indices $\varepsilon_0^{-1}P_{{\rm NL},i}=\sum_{jk}\chi_{ijk}^{(2)}E_jE_k+\sum_{jkl}\chi_{ijkl}^{(3)}E_jE_kE_l+\cdots$.
By assuming plane waves (no diffraction), we have $\nabla^2=\tfrac{\partial^2}{\partial z^2}$ so
\begin{align}\label{eq:pw-dz2}
\frac{\partial^2}{\partial z^2}\mathbf{E}(z,\omega)+ \frac{\omega^2}{c^2}\underline{\varepsilon}(\omega)\cdot\mathbf{E}(z,\omega) =-\frac{\omega^2}{c^2} \varepsilon_0^{-1}\mathbf{P}_{\rm NL}(z,\omega)
\end{align}
where $\underline{\varepsilon}(\omega)=1+\underline{\chi}^{(1)}(\omega)$ is the relative permittivity tensor, and we used $\varepsilon_0\mu_0=1/c^2$.

For SHG we define the electric field envelopes $\mathcal{E}_j$ for the  FW (frequency $\omega_1$) and the SH (frequency $\omega_2=2\omega_1$), as well as envelopes for the nonlinear polarization response $\mathcal{P}_{{\rm NL},j}$
\begin{align}\label{eq:E-env}
\mathbf{E}(z,t)&=\frac{1}{2}\left[\mathbf{u}_1 \mathcal{E}_1(z,t)e^{i k_1(\omega_1)z-i\omega_1 t} +\mathbf{u}_2 \mathcal{E}_2(z,t)e^{i k_2(\omega_2)z-i\omega_2 t}+c.c.\right]\\
\mathbf{P}_{\rm NL}(z,t)&=\frac{1}{2}\left[\mathbf{u}_1 \mathcal{P}_{\rm NL,1}(z,t)e^{i k_1(\omega_1)z-i\omega_1 t} +\mathbf{u}_2 \mathcal{P}_{\rm NL,2}(z,t)e^{i k_2(\omega_2)z-i\omega_2 t}+c.c.\right]
\label{eq:P-env}
\end{align}
where the wave numbers are defined as $k_j^2(\omega)=\tfrac{\omega^2}{c^2}\varepsilon_j(\omega)$ with $\varepsilon_j(\omega)=\mathbf{u}_j\cdot \underline{\chi}^{(1)}(\omega)$, and where $\mathbf{u}$ is a unit polarization vector. In a lossless medium $\varepsilon_j$ is real, and the linear refractive index is given by $n_j(\omega)=\sqrt{\varepsilon_j(\omega)}$.
The envelope wave equations are therefore
\begin{align}\label{eq:we-env-dz2}
    \frac{\partial^2\mathcal{E}_j(z,\omega)}{\partial z^2}+2ik_j(\omega_j) \frac{\partial \mathcal{E}_j(z,\omega) }{\partial z} +\left[k_j^2(\omega)-k_j^2(\omega_j)\right]\mathcal{E}_j(z,\omega) =-\frac{\omega^2}{c^2} \varepsilon_0^{-1}\mathcal{P}_{{\rm NL},j}(z,\omega).
\end{align}
The SVEA assumes $|\tfrac{\partial \mathcal{E}_j}{\partial z}|\ll k_j(\omega_j)|\mathcal{E}_j|$ giving
\begin{align}\label{eq:we-env}
i\frac{\partial \mathcal{E}_j(z,\omega) }{\partial z}+ \left[k_j(\omega)-k_j(\omega_j)\right]\mathcal{E}_j(z,\omega) =-\frac{\omega_j}{2n_j(\omega_j)c} \varepsilon_0^{-1}\mathcal{P}_{{\rm NL},j}(z,\omega)
\end{align}
where we used the approximation $k_j^2(\omega)-k_j^2(\omega_j)\simeq 2k_j(\omega_j)[k_j(\omega)-k_j(\omega_j)]$ \cite[p. 34]{agrawal:2007}. (A more accurate approach is to first go into the co-moving reference frame and then perform this expansion \cite[p. 14]{Couairon:2011}, but eventually the SVEA will discard terms so  the result will be the same.) At this time, we have also used that the assumption of a slow envelope compared to the fast oscillating carrier term $e^{-i\omega_j t}$ means that the nonlinear polarization term on the right-hand side can be approximated as $\tfrac{\omega^2}{2k_j(\omega_j) c^2} \varepsilon_0^{-1}\mathcal{P}_{{\rm NL},j}(\omega)\simeq \tfrac{\omega_j^2}{2k_j(\omega_j) c^2}\varepsilon_0^{-1} \mathcal{P}_{{\rm NL},j}(\omega)$. We can now perform the Taylor expansion of the wave number as $k_j(\omega)=\sum_{m=0}^\infty m!^{-1}k_j^{(m)}(\omega_m)(\omega-\omega_j)$, where the dispersion coefficients are defined as $k_j^{(m)}(\omega)={\rm d}^mk_j(\omega)/{\rm d}\omega^m$. Following this, we multiply both sides with $e^{-i\Omega t}$ where the frequency detuning is $\Omega=\omega-\omega_j$, and perform an inverse Fourier transform over $\Omega$ to time domain. This gives
\begin{align}\label{eq:we-env-time}
    \left[i\frac{\partial }{\partial z}+ik_j^{(1)}(\omega_j) \frac{\partial }{\partial t}
    +\hat D_j \right] \mathcal{E}_j(z,t) =-\frac{\omega_j}{2n_j(\omega_j)c} \varepsilon_0^{-1}\mathcal{P}_{{\rm NL},j}(z,t)
\end{align}
This is the SVEA equations in the stationary laboratory frame, describing a pulse envelope propagating with group velocity $1/k_j^{(1)}(\omega_j)$. The time-domain dispersion operator is
\begin{align}\label{eq:disp}
    \hat D_j=\sum_{m=2}^\infty k_j^{(m)}(\omega_j) \frac{i^m}{m!}\frac{\partial^m}{\partial t^m}
\end{align}
and it can be truncated at any order to study effects higher-order dispersion.

Besides a trivial transformation to a suitably chosen co-moving frame, what is left now is to evaluate the tensorial contributions to the nonlinear coefficients. Due to the vast variety of tensor combination possibilities for anisotropic nonlinearities it is here convenient to specify the problem very specifically: we therefore constrict ourselves to SHG in a uniaxial crystal, i.e. where light can be $o$-polarized with unit vector $\mathbf{e}^o$ or $e$-polarized with unit vector $\mathbf{e}^e$ \cite[Eq. (B6)]{bache:2010}, and we also restrict ourselves to degenerate SHG, where the two FW photons are indistinguishable. Let us define the quadratic nonlinear polarization response as $\mathbf{P}_{{\rm NL}}^{(2)}(z,t)=\underline{\chi}^{(2)}: \mathbf{E}(\mathbf{r},t)\mathbf{E}(\mathbf{r},t)$, where we have assumed that the quadratic nonlinearity is instantaneous in time. The procedure is then to insert the field envelope in this nonlinear polarization response and then calculate the polarization response for a specific field by applying the dot product with the fields unit vector: $\varepsilon_0^{-1}\mathbf{u}_j \cdot \mathbf{P}_{{\rm NL}}^{(2)}(z,t)$. The only relevant terms are those that oscillate with the fast carrier frequency oscillations of the field $e^{-i\omega_j t}$, as this defines the polarization envelope $\mathcal{P}$. Doing this we get for the FW and SH fields
\begin{align}\label{eq:PNL2}
    \varepsilon_0^{-1}\mathcal{P}^{(2)}_{{\rm NL},1}(z,t)=\chi^{(2)}_{\rm eff}\mathcal{E}_1^*(z,t)\mathcal{E}_2(z,t)e^{i\Delta k z}, \quad
    \varepsilon_0^{-1}\mathcal{P}^{(2)}_{{\rm NL},2}(z,t)=\chi^{(2)}_{\rm eff}\tfrac{1}{2}\mathcal{E}_1^2(z,t)e^{-i\Delta k z}
\end{align}
where the phase mismatch is $\Delta k=k_2(\omega_2)-2k_1(\omega_1)$. The effective susceptibility is calculated from the combination $\chi_{\rm eff}^{(2)}=\mathbf{u}_j\cdot \underline{\chi}^{(2)}:\mathbf{u}_k \mathbf{u}_l$ that appears from isolating the $e^{-i\omega_j t}$ term. The $\underline{\chi}^{(2)}$ is a rank 3 tensor with 27 different elements, but due to permutation symmetry in the the crystal axes indices there are only 18 nonzero elements \cite{boulanger:2006}. Depending on the crystal symmetry class (point group) many of these are zero. It is therefore fruitful to consider a specific point group, and perform the calculations. We show in Sec. \ref{sec:anisotropy} the results for the point group $3m$. We note that the nonlinear polarizations would have been a factor of 2 larger had we not used a factor $\tfrac{1}{2}$ in front of our envelope definition.

Moving to the cubic nonlinear tensor component, it is done in much the same way, and we refer to \cite[App. B]{bache:2010} for details. We again consider an instantaneous response, and the cubic nonlinear polarization response that oscillates with $e^{-i\omega_j t}$ turns out to be related to SPM and XPM; we could have had contributions from third-harmonic generation here if we had included such a harmonic field $\mathcal{E}_3(z,t)e^{i k_3(\omega_3)z-i\omega_3 t}$ in the envelope definition. We then get \cite[Eq. (B8)]{bache:2010}
\begin{align}\label{eq:PNL3}
    \varepsilon_0^{-1}\mathcal{P}^{(3)}_{{\rm NL},j}(z,t)&=\frac{1}{4}\left[
    3\chi^{(3)}_{{\rm eff}}(\omega_j;\omega_j)|\mathcal{E}_j(z,t)|^2
    +6\chi^{(3)}_{{\rm eff}}(\omega_j;\omega_k)|\mathcal{E}_k(z,t)|^2\right]\mathcal{E}_j(z,t)
\end{align}
The notation of the cubic susceptibilities $\chi^{(3)}_{\rm eff}(\omega_i;\omega_j)$ implicitly assumes phase-matched SPM and XPM interaction between $\omega_i$ and $\omega_j$, as opposed to general four-wave mixing. We always have $\chi^{(3)}_{\rm eff}(\omega_1;\omega_2)=\chi^{(3)}_{\rm eff}(\omega_2;\omega_1)$. The factor $\tfrac{1}{4}$ comes from the factor $\tfrac{1}{2}$ in front of our envelope definitions, Eqs. (\ref{eq:E-env})-(\ref{eq:P-env}); if we do not have this factor $\tfrac{1}{2}$, as in e.g. Boyd \cite{boyd:2007}, it is important to note that the $\chi^{(3)}_{{\rm eff}}$ values (and consequently also the $\nk$-values) in the two cases remain the same. The $\underline{\chi}^{(3)}$ is a rank 4 tensor with 81 different elements, but due to permutation symmetry in the the crystal axes indices there are only 30 nonzero elements \cite{boulanger:2006}. The calculations of the $\chi^{(3)}_{{\rm eff}}$ specific to a uniaxial crystal in the $3m$ point group was done in \cite[App. B]{bache:2010}, and in Sec. \ref{sec:anisotropy} we summarize the results.

We can now express the plane-wave SVEA equations for the electric field envelopes as
\begin{align}
\left[  i\frac{\partial}{\partial z}+ \hat D_1
\right ] \mathcal{E}_1+\frac{\omega_1 d_{\rm eff}}{n_1 c} \mathcal{E}_1^* \mathcal{E}_2 e^{i\Delta k z}
\nonumber\\
+\frac{3\omega_1}{8n_1c}\left[\chi_{\rm eff}^{(3)}(\omega_1;\omega_1)\mathcal{E}_1|\mathcal{E}_1|^2 +  2\chi_{\rm eff}^{(3)}(\omega_1;\omega_2)\mathcal{E}_1|\mathcal{E}_2|^2 \right] =0
\label{eq:FW}
\\
\left[  i\frac{\partial}{\partial z}
-id_{12}\frac{\partial}{\partial    \tau}
+\hat D_2
\right ] \mathcal{E}_2 +\frac{\omega_2 d_{\rm eff}}{n_2 c}\tfrac{1}{2}\mathcal{E}_1^2 e^{-i\Delta k z}
\nonumber\\
+\frac{3\omega_2}{8n_2c}\left[\chi_{\rm eff}^{(3)}(\omega_2;\omega_2)\mathcal{E}_2|\mathcal{E}_2|^2 +  2\chi_{\rm eff}^{(3)}(\omega_2;\omega_1)\mathcal{E}_2|\mathcal{E}_1|^2 \right]=0
\label{eq:SH}
\end{align}
The time $\tau$ follows the FW group-velocity $1/\kp_1(\omega_1)$ by the transformation from the lab time $t$ as $\tau=t-z\kp_1(\omega_1)$, which gives the group-velocity mismatch (GVM) term  $d_{12}=\kp_1(\omega_1)-\kp_2(\omega_2)$. We use the short-hand notation $n_j\equiv n_j(\omega_j)$, and $d_{\rm eff}=\chi^{(2)}_{\rm eff}/2$ is the usual reduced notation of the effective quadratic nonlinearity. We can convert the electric field to intensity $\mathcal{E}_j\rightarrow (2/\varepsilon_0 n_j c)^{1/2}A_j$, so $|A_j|^2$ is the intensity in $[{\rm W/m^2}]$, and we get
\begin{align}
\left[  i\frac{\partial}{\partial z}+\hat D_1
\right ] A_1+ \kappa_{\rm SHG}^I A_1^* A_2 e^{i\Delta k z}
\nonumber\\
+\frac{\omega_1}{c}\left[\nk(\omega_1;\omega_1)A_1|A_1|^2 +  2 \nk(\omega_1;\omega_2)A_1|A_2|^2 \right] =0
\label{eq:FW-int}
\\
\left[  i\frac{\partial}{\partial z}
-id_{12}\frac{\partial}{\partial    \tau}
+\hat D_2
\right] A_2 +\kappa_{\rm SHG}^I  A_1^2  e^{-i\Delta k z}
\nonumber\\
+\frac{\omega_2}{c}\left[\nk(\omega_2;\omega_2)A_2|A_2|^2 +  2\nk(\omega_1;\omega_2)A_2|A_1|^2 \right]=0
\label{eq:SH-int}
\end{align}
where the equations now have a common SHG nonlinear parameter
\begin{align}\label{eq:kappa_SHG}
\kappa_{\rm SHG}^I=\frac{\omega_1 d_{\rm eff}}{n_1 c}\sqrt{\frac{2}{n_2 \varepsilon_0 c}}
\end{align}
We can also establish the link between the Kerr nonlinear refractive indices and the cubic nonlinear susceptibilities as 
\begin{align}\label{eq:n2-chi3}
 \nk(\omega_i;\omega_j)=   \frac{3 \chi_{\rm eff}^{(3)}(\omega_i;\omega_j)} {4n_in_j\varepsilon_0 c}
\end{align}
cf. also Eq. (C6) in \cite{bache:2010}. Both $\nk$ and $\chi^{(3)}$ are in mks (SI) units, while conversion to and from the esu system is reported in \cite{bache:2010}. 

The SVEA equations can be extended to the slowly-evolving wave equations (SEWA) \cite{moses:2006,bache:2007} by including self-steepening effects, and also the non-instantaneous (delayed) Raman effect can straightforwardly be included \cite{bache:2007,guo:2012}. For simplicity such effects are neglected here.

\section{Reduced nonlinear Schr{\"o}dinger equation in the strong cascading limit}
\label{sec:NLSE}

The cascading nonlinear contribution in the strong cascading limit ($\Delta k L\gg 2\pi$) can quickly be found by taking assuming an undepleted FW. Thus $E_1$ is taken independent on $z$ in Eq. (\ref{eq:SH}), and the SH can therefore be solved directly when the Kerr SPM and XPM terms are neglected (which follows directly from the undepleted FW assumption). In absence of dispersion (the effect of dispersion is discussed in Sec. \ref{sec:implications}) we get that the SH is slaved to the FW: $\mathcal{E}_{2,\rm casc}(z,\tau)=-\Delta k^{-1} \tfrac{\omega_2 d_{\rm eff}} {n_2 c} \frac{1}{2}\mathcal{E}_1^2(z,\tau) e^{-i\Delta k z}$. When inserting this in the FW equation (\ref{eq:FW}) we arrive at the following nonlinear Schr{\"o}dinger equation for the FW
\begin{align}\label{eq:NLSE}
\left[  i\frac{\partial}{\partial z}+\hat D_1
\right ] \mathcal{E}_1
+\frac{3\omega_1}{8n_1c}\left[\chi_{\rm casc}^{(3)}+\chi_{\rm eff}^{(3)}(\omega_1;\omega_1) \right]\mathcal{E}_1|\mathcal{E}_1|^2
+\frac{5\omega_1}{16 n_1 c}\chi_{\rm casc}^{(5)}\mathcal{E}_1|\mathcal{E}_1|^4  =0
\end{align}
From these equations we now understand the basis of the nonlinear index change described by Eq. (\ref{eq:Deltan}): in the strong cascading limit the FW experiences a cubic Kerr self-action nonlinearity $\propto |\mathcal{E}_1|^2$, where both cascading and material Kerr nonlinearities contribute. The cascading Kerr-like nonlinearity is described by the parameters
\begin{align}
\label{eq:chi3-SHG}
  \chi_{\rm casc}^{(3)}&=-\frac{8\omega_1 d_{\rm  eff}^2}{3cn_2 \Delta k}
,\quad
  \ns=-\frac{2\omega_1 d_{\rm  eff}^2}{c^2\varepsilon_0 n_1^2n_2 \Delta k}
\end{align}
where the latter is identical to Eq. (\ref{eq:n2-SHG}) and it comes from using Eq. (\ref{eq:n2-chi3}), or simply performing the same exercise with the intensity version of the SVEA equations. These will be used in Sec. \ref{sec:cubic} to estimate the cascading contribution in the various experiments. A practical note: this lowest order cascading nonlinearity seemingly diverges at $\Delta k=0$, but this is because the assumption of an undepleted FW breaks down in this limit. To the next order a correction term $1-{\rm sinc} (\Delta k L)$ must be applied \cite{bache:2012-HOKE}, which resolves the divergence at $\Delta k=0$ and from this the maximum cascading strength is obtained at $\Delta kL\simeq \pi$.

There is also a quintic nonlinearity from cascading mixed with the XPM term of the FW
\begin{align}\label{eq:chi5casc}
\chi_{\rm casc}^{(5)}&=\frac{12 \omega_1^2 \chi_{\rm eff}^{(3)}(\omega_1;\omega_2) d_{\rm eff}^2} {5 n_2^2 c^2 \Delta k^2},
\quad     n_{4,\rm casc}^I=\frac{4 \omega_1^2 n_{\rm Kerr}^{I}(\omega_1;\omega_2) d_{\rm eff}^2}{\Delta k^2 n_1^2 n_2 \varepsilon_0 c^3}
\end{align}
where the latter comes from the relationship $n_{4}^I=5\chi^{(5)}/(4 n_1^3\varepsilon_0^2 c^2)$, and the quintic nonlinear refractive index $n_4^I$ was defined in Eq. (\ref{eq:Deltan_n2}). 
This term gives a nonlinear index change that is positive and $\propto I^2$, and becomes important for high pulse fluences \cite{bache:2007,zeng:2012}. However, we evaluate it to be negligible in the experiments treated in Sec. \ref{sec:cubic}.

Note that these simple cascading expressions only hold in the strong cascading limit, which essentially means that the characteristic length for the phase mismatch (i.e. the coherence length) is much shorter than any other characteristic length scale in the system \cite{bache:2007a}. This can be achieved by fulfillment of two criteria: (a) a strong phase mismatch $\Delta kL\gg 2\pi$ (i.e. the many up- and down-conversion steps must occur inside the interaction length). (b) A phase mismatch that dominates over the quadratic nonlinearity. To express this
we introduce the traditional \cite{desalvo:1992} quadratic nonlinear strength parameter
$\Gamma=\omega_1 d_{\rm eff}|\mathcal{E}_{\rm in}|/(\sqrt{n_1 n_2} c)$,
where $|\mathcal{E}_{\rm in}|$ is the peak input electric field. Criterion (b) can then be expressed as $\Delta k\gg \Gamma$ (which essentially ensures an undepleted FW, see also the discussion in \cite{bache:2012-HOKE}). In all the cases in Sec. \ref{sec:cubic} where we correct for cascading contributions, we evaluate that the strong cascading limit is fulfilled.

\section{BBO crystal parameters}
\label{sec:Crystal-parameters}


The crystal beta-barium-borate ($\beta$-{BaB}$_2${O}$_4$, BBO) is a negative uniaxial crystal of the point group $3m$, which has a transmission range from 189-3500 nm \cite{chen:1985,Nikogosyan:1991}. The Sellmeier equations were taken from Zhang et al. \cite{Zhang:2000}. We used the quadratic nonlinear tensor components measured by Shoji et al. \cite{Shoji:1999}: at $\lambda=1064$ nm $|d_{22}|=2.2$ pm/V, $|d_{31}|=|d_{22}| 0.018=0.04$ pm/V, $d_{15}=0.03$ pm/V, and $d_{33}=0.04$ pm/V; at $\lambda=852$ nm $|d_{22}|=2.3$ pm/V; at $\lambda=532$ nm $|d_{22}|=2.6$ pm/V; at $\lambda=1313$ nm $|d_{22}|=1.9$ pm/V. Note that Kleinman symmetry has $d_{31}=d_{15}$, which is only slightly violated at 1064 nm by these measurements; we therefore throughout this paper assume Kleinman symmetry. The sign of the product $d_{31} d_{22}$ has been shown to be negative \cite{klein:2003}, which means that flipping the crystal 180$^\circ$ gives a changed effective nonlinearity (see also discussion for Fig. \ref{fig:cut}). An overview of other measurements is given in \cite{Eckardt:2004}; in particular note that the 1064 nm $d_{22}$ value is similar in most measurements, but the $d_{15}$ value was historically measured to be higher. This stems from the initial measurements of BBO \cite{chen:1985} that gave $d_{15}=0.07d_{22}$ \cite{Nikogosyan:1991}, which then with the $d_{22}=2.2$ pm/V value gives $d_{15}\simeq0.15$ pm/V. A users note of the Shoji et al. values: the 852 and 1064 nm measurements of the $|d_{22}|$ value agree well with Miller's scaling (they have the same Miller's delta \cite{Shoji:1999}), while the 1313 nm measurement seems to lie too low. Therefore when calculating the effective nonlinearity at an arbitrary wavelength the safest option generally is to use the 1064 nm values, where all the nonlinear components were measured, and employ Miller's scaling to go to other wavelengths.



\begin{thebibliography}{10}
\newcommand{\enquote}[1]{``#1''}

\bibitem{Ostrovskii:1967}
L.~A. Ostrovskii, \enquote{Self-action of light in crystals,} Pis'ma Zh. Eksp.
  Teor. Fiz. \textbf{5}, 331 (1967). [JETP Lett. {\bf 5}, 272-275 (1967)].

\bibitem{Thomas:1972}
J.~M.~R. Thomas and J.~P.~E. Taran, \enquote{Pulse distortions in mismatched
  second harmonic generation,} Opt. Commun. \textbf{4}, 329 -- 334 (1972).

\bibitem{desalvo:1992}
R.~DeSalvo, D.~Hagan, M.~{Sheik-Bahae}, G.~Stegeman, E.~W. {Van Stryland}, and
  H.~Vanherzeele, \enquote{Self-focusing and self-defocusing by cascaded
  second-order effects in {KTP},} Opt. Lett. \textbf{17}, 28--30 (1992).

\bibitem{stegeman:1996}
G.~I. Stegeman, D.~J. Hagan, and L.~Torner, \enquote{$\chi^{(2)}$ cascading
  phenomena and their applications to all-optical signal processing,
  mode-locking, pulse compression and solitons,} Opt. Quantum Electron.
  \textbf{28}, 1691--1740 (1996).

\bibitem{liu:1999}
X.~Liu, L.-J. Qian, and F.~W. Wise, \enquote{High-energy pulse compression by
  use of negative phase shifts produced by the cascaded $\chi^{(2)}:\chi^{(2)}$
  nonlinearity,} Opt. Lett. \textbf{24}, 1777--1779 (1999).

\bibitem{Beckwitt:2001}
K.~Beckwitt, F.~W. Wise, L.~Qian, L.~A. {Walker II}, and E.~Canto-Said,
  \enquote{Compensation for self-focusing by use of cascade quadratic
  nonlinearity,} Opt. Lett. \textbf{26}, 1696--1698 (2001).

\bibitem{ashihara:2002}
S.~Ashihara, J.~Nishina, T.~Shimura, and K.~Kuroda, \enquote{Soliton
  compression of femtosecond pulses in quadratic media,} J. Opt. Soc. Am. B
  \textbf{19}, 2505--2510 (2002).

\bibitem{ilday:2004}
F.~{\" O}. Ilday, K.~Beckwitt, Y.-F. Chen, H.~Lim, and F.~W. Wise,
  \enquote{Controllable {R}aman-like nonlinearities from nonstationary,
  cascaded quadratic processes,} J. Opt. Soc. Am. B \textbf{21}, 376--383
  (2004).

\bibitem{moses:2006}
J.~Moses and F.~W. Wise, \enquote{Soliton compression in quadratic media:
  high-energy few-cycle pulses with a frequency-doubling crystal,} Opt. Lett.
  \textbf{31}, 1881--1883 (2006).

\bibitem{moses:2006b}
J.~Moses and F.~W. Wise, \enquote{Controllable self-steepening of ultrashort
  pulses in quadratic nonlinear media,} Phys. Rev. Lett. \textbf{97}, 073903
  (2006).

\bibitem{moses:2007}
J.~Moses, E.~Alhammali, J.~M. Eichenholz, and F.~W. Wise, \enquote{Efficient
  high-energy femtosecond pulse compression in quadratic media with flattop
  beams,} Opt. Lett. \textbf{32}, 2469--2471 (2007).

\bibitem{bache:2010e}
M.~Bache, O.~Bang, B.~B. Zhou, J.~Moses, and F.~W. Wise, \enquote{Optical
  {C}herenkov radiation in ultrafast cascaded second-harmonic generation,}
  Phys. Rev. A \textbf{82}, 063806 (2010).

\bibitem{tan:1993}
H.~Tan, G.~P. Banfi, and A.~Tomaselli, \enquote{Optical frequency mixing
  through cascaded second-order processes in beta-barium borate,} Appl. Phys.
  Lett. \textbf{63}, 2472--2474 (1993).

\bibitem{hache:1995}
F.~Hache, A.~Z{\'e}boulon, G.~Gallot, and G.~M. Gale, \enquote{Cascaded
  second-order effects in the femtosecond regime in $\beta$-barium borate:
  self-compression in a visible femtosecond optical parametric oscillator,}
  Opt. Lett. \textbf{20}, 1556--1558 (1995).

\bibitem{moses:2007a}
J.~Moses, B.~A. Malomed, and F.~W. Wise, \enquote{Self-steepening of ultrashort
  optical pulses without self-phase modulation,} Phys. Rev. A \textbf{76},
  021802(R) (2007).

\bibitem{desalvo:1996}
R.~DeSalvo, A.~A. Said, D.~Hagan, E.~W. {Van Stryland}, and M.~{Sheik-Bahae},
  \enquote{Infrared to ultraviolet measurements of two-photon absorption and
  n$_2$ in wide bandgap solids,} IEEE J. Quantum Electron. \textbf{32},
  1324--1333 (1996).

\bibitem{sheik-bahae:1997}
M.~Sheik-Bahae and M.~Ebrahimzadeh, \enquote{Measurements of nonlinear
  refraction in the second-order $\chi^{(2)}$ materials {KTiOPO}$_4$,
  {KNbO}$_3$, $\beta$-{BaB}$_2${O}$_4$ and {LiB}$_3${O}$_5$,} Opt. Commun.
  \textbf{142}, 294--298 (1997).

\bibitem{li:1997}
H.~Li, F.~Zhou, X.~Zhang, and W.~Ji, \enquote{Bound electronic kerr effect and
  self-focusing induced damage in second-harmonic-generation crystals,} Opt.
  Commun. \textbf{144}, 75 -- 81 (1997).

\bibitem{Li:2001}
H.~P. Li, C.~H. Kam, Y.~L. Lam, and W.~Ji, \enquote{Femtosecond {Z}-scan
  measurements of nonlinear refraction in nonlinear optical crystals,} Opt.
  Mater. \textbf{15}, 237--242 (2001).

\bibitem{ganeev:2003}
R.~Ganeev, I.~Kulagin, A.~Ryasnyanskii, R.~Tugushev, and T.~Usmanov,
  \enquote{The nonlinear refractive indices and nonlinear third-order
  susceptibilities of quadratic crystals,} Opt. Spectrosc. \textbf{94},
  561--568 (2003). [Opt. Spektrosk. {\bf 94}, 615-623 (2003)].

\bibitem{Banks:2002}
P.~S. Banks, M.~D. Feit, and M.~D. Perry, \enquote{High-intensity
  third-harmonic generation,} J. Opt. Soc. Am. B \textbf{19}, 102--118 (2002).

\bibitem{Midwinter:1965}
J.~E. Midwinter and J.~Warner, \enquote{The effects of phase matching method
  and of crystal symmetry on the polar dependence of third-order non-linear
  optical polarization,} Br. J. Appl. Phys. \textbf{16}, 1667--1674 (1965).

\bibitem{wang:1969}
C.~Wang and E.~Baardsen, \enquote{Optical third harmonic generation using
  mode-locked and non-mode-locked lasers,} Appl. Phys. Lett. \textbf{15},
  396--–397 (1969).

\bibitem{Sheik-Bahae:1991}
M.~Sheik-Bahae, D.~Hutchings, D.~Hagan, and E.~Van~Stryland,
  \enquote{Dispersion of bound electron nonlinear refraction in solids,} IEEE
  J. Quantum Electron. \textbf{27}, 1296 --1309 (1991).

\bibitem{ashihara:2004}
S.~Ashihara, T.~Shimura, K.~Kuroda, N.~E. Yu, S.~Kurimura, K.~Kitamura, M.~Cha,
  and T.~Taira, \enquote{Optical pulse compression using cascaded quadratic
  nonlinearities in periodically poled lithium niobate,} Appl. Phys. Lett.
  \textbf{84}, 1055--1057 (2004).

\bibitem{Zeng:2008}
X.~Zeng, S.~Ashihara, X.~Chen, T.~Shimura, and K.~Kuroda, \enquote{Two-color
  pulse compression in aperiodically-poled lithium niobate,} Opt. Commun.
  \textbf{281}, 4499 -- 4503 (2008).

\bibitem{zhou:2012}
B.~B. Zhou, A.~Chong, F.~W. Wise, and M.~Bache, \enquote{Ultrafast and
  octave-spanning optical nonlinearities from strongly phase-mismatched
  quadratic interactions,} Phys. Rev. Lett. \textbf{109}, 043902 (2012).


\bibitem{boyd:2007}
R.~W. Boyd, \emph{Nonlinear Optics} (Academic Press, 2007), 3rd ed.

\bibitem{Nikogosyan:1991}
D.~N. Nikogosyan, \enquote{Beta barium borate ({BBO}) - {A} review of its
  properties and applications,} Appl. Phys. A \textbf{52}, 359--368 (1991).

\bibitem{bache:2010}
M.~Bache and F.~W. Wise, \enquote{Type-{I} cascaded quadratic soliton
  compression in lithium niobate: Compressing femtosecond pulses from
  high-power fiber lasers,} Phys. Rev. A \textbf{81}, 053815 (2010).

\bibitem{Langrock:2007}
C.~Langrock, M.~M. Fejer, I.~Hartl, and M.~E. Fermann, \enquote{Generation of
  octave-spanning spectra inside reverse-proton-exchanged periodically poled
  lithium niobate waveguides,} Opt. Lett. \textbf{32}, 2478--2480 (2007).

\bibitem{Phillips:2011}
C.~R. Phillips, C.~Langrock, J.~S. Pelc, M.~M. Fejer, I.~Hartl, and M.~E.
  Fermann, \enquote{Supercontinuum generation in quasi-phasematched
  waveguides,} Opt. Express \textbf{19}, 18754--18773 (2011).

\bibitem{boulanger:2006}
B.~Boulanger and J.~Zyss, \emph{International Tables for Crystallography}
  (Springer, 2006), vol. D: Physical Properties of Crystals, chap. 1.7
  Nonlinear optical properties, pp. 178--219.

\bibitem{miller:1964}
R.~C. Miller, \enquote{Optical second harmonic generation in piezoelectric
  crystals,} Appl. Phys. Lett. \textbf{5}, 17--19 (1964).

\bibitem{wynne:1969}
J.~J. Wynne, \enquote{Optical third-order mixing in {GaAs}, {Ge}, {Si}, and
  {InAs},} Phys. Rev. \textbf{178}, 1295--1303 (1969).

\bibitem{Ettoumi:2010}
W.~Ettoumi, Y.~Petit, J.~Kasparian, and J.-P. Wolf, \enquote{Generalized
  {M}iller formul{\ae},} Opt. Express \textbf{18}, 6613--6620 (2010).

\bibitem{bache:2007}
M.~Bache, J.~Moses, and F.~W. Wise, \enquote{Scaling laws for soliton pulse
  compression by cascaded quadratic nonlinearities,} J. Opt. Soc. Am. B
  \textbf{24}, 2752--2762 (2007).

\bibitem{sheik-bahae:1990}
M.~Sheik-Bahae, A.~Said, T.-H. Wei, D.~Hagan, and E.~Van~Stryland,
  \enquote{Sensitive measurement of optical nonlinearities using a single
  beam,} IEEE J. Quantum Electron. \textbf{26}, 760--769 (1990).

\bibitem{krauss:1994}
T.~D. Krauss and F.~W. Wise, \enquote{Femtosecond measurement of nonlinear
  absorption and refraction in {CdS}, {ZnSe}, and {ZnS},} Appl. Phys. Lett.
  \textbf{65}, 1739--1741 (1994).

\bibitem{Gnoli:2005}
A.~Gnoli, L.~Razzari, and M.~Righini, \enquote{Z-scan measurements using high
  repetition rate lasers: how to manage thermal effects,} Opt. Express
  \textbf{13}, 7976--7981 (2005).

\bibitem{Nibbering:1995}
E.~Nibbering, M.~Franco, B.~Prade, G.~Grillon, C.~L. Blanc, and A.~Mysyrowicz,
  \enquote{Measurement of the nonlinear refractive index of transparent
  materials by spectral analysis after nonlinear propagation,} Opt. Commun.
  \textbf{119}, 479 -- 484 (1995).

\bibitem{Shoji:1999}
I.~Shoji, H.~Nakamura, K.~Ohdaira, T.~Kondo, R.~Ito, T.~Okamoto, K.~Tatsuki,
  and S.~Kubota, \enquote{Absolute measurement of second-order
  nonlinear-optical coefficients of $\beta$-{B}a{B}$_2${O}$_4$ for visible to
  ultraviolet second-harmonic wavelengths,} J. Opt. Soc. Am. B \textbf{16},
  620--624 (1999).

\bibitem{Zhang:2000}
D.~Zhang, Y.~Kong, and J.~Zhang, \enquote{Optical parametric properties of
  532-nm-pumped beta-barium-borate near the infrared absorption edge,} Opt.
  Commun. \textbf{184}, 485--491 (2000).

\bibitem{Ganeev:2012}
R.~A. Ganeev (2012). Private communication.

\bibitem{moses:2010comm}
J.~A. Moses (2010). Private communication.

\bibitem{guo:2012}
H.~Guo, X.~Zeng, B.~Zhou, and M.~Bache, \enquote{Electric field modeling and
  self-steepening counterbalance of cascading nonlinear soliton pulse
  compression,} J. Opt. Soc. Am. B  (2012). Submitted, arXiv:1210.5903.

\bibitem{bosshard:2000}
C.~Bosshard, U.~Gubler, P.~Kaatz, W.~Mazerant, and U.~Meier,
  \enquote{Non-phase-matched optical third-harmonic generation in
  noncentrosymmetric media: Cascaded second-order contributions for the
  calibration of third-order nonlinearities,} Phys. Rev. B \textbf{61},
  10688--10701 (2000).

\bibitem{Sheik-Bahae:1997-KLAC}
M.~Sheik-Bahae, \enquote{Femtosecond kerr-lens autocorrelation,} Opt. Lett.
  \textbf{22}, 399--401 (1997).

\bibitem{fan:1989}
Y.~Fan, R.~Eckardt, R.~Byer, C.~Chen, and A.~Jiang, \enquote{Barium borate
  optical parametric oscillator,} IEEE J. Quantum Elec. \textbf{25}, 1196
  --1199 (1989).

\bibitem{Boling:1978}
N.~Boling, A.~Glass, and A.~Owyoung, \enquote{Empirical relationships for
  predicting nonlinear refractive index changes in optical solids,} IEEE J.
  Quantum Electron. \textbf{14}, 601 -- 608 (1978).

\bibitem{bache:2007a}
M.~Bache, O.~Bang, J.~Moses, and F.~W. Wise, \enquote{Nonlocal explanation of
  stationary and nonstationary regimes in cascaded soliton pulse compression,}
  Opt. Lett. \textbf{32}, 2490--2492 (2007).

\bibitem{bache:2008}
M.~Bache, O.~Bang, W.~Krolikowski, J.~Moses, and F.~W. Wise, \enquote{Limits to
  compression with cascaded quadratic soliton compressors,} Opt. Express
  \textbf{16}, 3273--3287 (2008).

\bibitem{zeng:2012}
X.~Zeng, H.~Guo, B.~Zhou, and M.~Bache, \enquote{Soliton compression to
  few-cycle pulses with a high quality factor by engineering cascaded quadratic
  nonlinearities,} Opt. Express \textbf{20}, 27071--27082 (2012).
  ArXiv:1210.5928.

\bibitem{agrawal:2007}
G.~P. Agrawal, \emph{Nonlinear fiber optics} (Academic Press, San Diego, 2007),
  4th ed.

\bibitem{Couairon:2011}
A.~Couairon, E.~Brambilla, T.~Corti, D.~Majus, O.~{de J.
  Ram\'{\i}rez-G\'{o}ngora}, and M.~Kolesik, \enquote{Practitioner's guide to
  laser pulse propagation models and simulation,} Eur. Phys. J. Special Topics
  \textbf{199}, 5--76 (2011).

\bibitem{bache:2012-HOKE}
M.~Bache, F.~Eilenberger, and S.~Minardi, \enquote{Higher-order {K}err effect
  and harmonic cascading in gases,} Opt. Lett. \textbf{37}, 4612--4614 (2012).

\bibitem{chen:1985}
C.~Chen, B.~Wu, A.~Jiang, and G.~You, \enquote{A new-type ultraviolet {SHG}
  crystal - beta-{BaB$_2$O$_4$},} Sci. Sin., Ser. B \textbf{28}, 235--243
  (1985).

\bibitem{klein:2003}
R.~S. Klein, G.~E. Kugel, A.~Maillard, A.~Sifi, and K.~Polgar,
  \enquote{Absolute non-linear optical coefficients measurements of {BBO}
  single crystal and determination of angular acceptance by second harmonic
  generation,} Opt. Mater. \textbf{22}, 163--169 (2003).

\bibitem{Eckardt:2004}
R.~C. Eckardt and G.~C. Cattela, \enquote{Characterization techniques for
  second-order nonlinear optical materials,} Proc. SPIE \textbf{5337}, 1--10
  (2004).

\end{thebibliography}
\end{document}